# Highlights

## Fundamental comparison between the pseudopotential and the free energy lattice Boltzmann methods

Luiz Eduardo Czelusniak, Ivan Talão Martins, Luben Cabezas Gómez, Natan Augusto Vieira Bulgarelli, William Monte Verde, Marcelo Souza de Castro

- Proposition of modified pseudopotential LBM for tuning interface width independently of EOS selection.
- Presentation of fundamental comparison between pseudopotential and free energy multiphase LBM.
- The free energy model was more accurate than the pseudopotential model for the same reduced temperatures
- The pseudopotential model enabled the simulation of lower reduced temperatures

# Fundamental comparison between the pseudopotential and the free energy lattice Boltzmann methods

Luiz Eduardo Czelusniak[a,b,*], Ivan Talão Martins[c], Luben Cabezas Gómez[c], Natan Augusto Vieira Bulgarelli[a], William Monte Verde[a] and Marcelo Souza de Castro[a]

[a]*Center for Energy and Petroleum Studies, State University of Campinas, Campinas, 13083-896, São Paulo, Brazil*
[b]*Institute for Applied and Numerical Mathematics, Karlsruhe, D-76131, Baden-Württemberg, Germany*
[c]*Department of Mechanical Engineering, São Carlos School of Engineering, University of São Paulo, São Carlos, 13566-590, São Paulo, Brazil*

ABSTRACT

The pseudopotential and free energy models are two popular extensions of the lattice Boltzmann method for multiphase flows. Until now, they have been developed apart from each other in the literature. However, important questions about whether each method performs better needs to be solved. In this work, we perform a fundamental comparison between both methods through basic numerical tests. This comparison is only possible because we developed a novel approach for controlling the interface thickness in the pseudopotential method independently on the equation of state. In this way, it is possible to compare both methods maintaining the same equilibrium densities, interface thickness, surface tension and equation of state parameters. The well-balanced approach was selected to represent the free energy. We found that the free energy one is more practical to use, as it is not necessary to carry out previous simulations to determine simulation parameters (interface thickness, surface tension, etc). In addition, the tests proofed that the free energy model is more accurate than the pseudopotential model. Furthermore, the pseudopotential method suffers from a lack of thermodynamic consistency even when applying the corrections proposed in the literature. On the other hand, for both static and dynamic tests we verified that the pseudopotential method was able to simulate lower reduced temperature than the free energy one. We hope that these results will guide authors in the use of each method.

## 1. INTRODUCTION

The lattice Boltzmann method (LBM) has growing as a powerful tool for multiphase simulations, considering or not the phase-change process. Inside the LBM framework for multiphase single-component fluids, there are two well-known extensions: the pseudopotential method (Shan and Chen, 1993) (P-LBM) and the free-energy method (Swift, Orlandini, Osborn and Yeomans, 1996) (FE-LBM).

The first versions of the P-LBM had some issues concerning thermodynamic consistency and surface tension adjustment. Along years authors proposed several improvements, such as improved thermodynamic consistency (Li, Luo and Li, 2013; Peng, Ayala, Wang and Ayala, 2020), surface tension adjustment (Li and Luo, 2013), higher-order error correction (Lycett-Brown and Luo, 2015; Huang and Wu, 2016; Wu, Gui, Yang, Tu and Jiang, 2018), multirange interactions (Sbragaglia, Benzi, Biferale, Succi, Sugiyama and Toschi, 2007) and understanding about the spurious currents (Shan, 2006; Peng, Ayala, Ayala and Wang, 2019). Afterward, P-LBM became widely employed in several multiphase applications, as for example pool-boiling simulations (Li, Kang, Francois, He and Luo, 2015; Fei, Yang, Chen, Mo and Luo, 2020), heated channel flows (Wang, Lou and Li, 2020; Zhang, Chen, Ji, Liu, Liu and Tao, 2021) or to study the influence of micro-structured surfaces during phase-change (Li, Li, Yu and Luo, 2021; Wang, Liang, Yin and Shen, 2023).

The applicability of the second method (FE-LBM) faced also several challenges with previous schemes, suffering from lack of thermodynamic consistency and galilean invariance (Kikkinides, Yiotis, Kainourgiakis and Stubos, 2008). Authors proposed corrections that worked for 1D (one-dimensional) cases (Wagner, 2006), for temperatures close to the critical one (Kikkinides et al., 2008), and using lattices with more velocities than D2Q9 (Siebert, Philippi and

---

Fundamental comparison between the pseudopotential and the free energy lattice Boltzmann methods

Mattila, 2014) for 2D cases, for example. Recently, Guo (2021) introduced the well-balanced free energy method (WB-FE-LBM). This approach is very promising because it is able to guarantee thermodynamic consistency for 2D cases using regular D2Q9 lattice.

So far, literary research on both methods has been carried out separately (Sudhakar and Das, 2020). However, it holds great scientific interest to understand how one can compare these two methods to each other. Therefore, this understanding will provide for LBM users which one is the most adequate method to use, depending on the problem they aim to solve. Thereby, the goal of this work is to carry out a theoretical comparative study between the P-LBM and FE-LBM methods. Due to the lack of similar studies in the literature, this research was carried out focused on more fundamental problems. The results are verified with basic numerical simulations. In future works, we will explore how the two methods behave in relation to more complex problems. For our purpose, we will investigate mainly aspects related to accuracy and physical consistency.

The article is organized as follows: Section. 2 summarizes the LBM, P-LBM and FE-LBM. In Sec. 3 the methodology used to determine the simulation parameters and the equation of state used in this work are presented. In Sec. 4 we present a procedure to change the P-LBM interface thickness without changing the equation of state parameters. This procedure will be necessary for the fair comparison between P-LBM and FE-LBM. In Sec. 5 we present the benchmark simulation results comparing both methods, which involve static and dynamic tests. Finally, in Sec. 6, a conclusion to the work is presented.

## 2. THEORETICAL BACKGROUND

The lattice Boltzmann equation (LBE) with the multiple-relaxation time (MRT) collision operator can be written as

$$f_i(t + \Delta t, \mathbf{x} + \mathbf{c}_i \Delta t) - f_i(t, \mathbf{x}) = -\Delta t \left[ \mathbf{M}^{-1} \mathbf{\Lambda} \mathbf{M} \right]_{ij} (f_j - f_j^{eq}) + \Delta t \left[ \mathbf{M}^{-1} \left( \mathbf{I} - \frac{\mathbf{\Lambda} \Delta t}{2} \right) \mathbf{M} \right]_{ij} F'_j + \Delta t C_i, \quad (1)$$

where $f_i$ are the particle distribution functions related to the particle velocities $\mathbf{c}_i$ and $f_i^{eq}$ are the local equilibrium distribution functions (EDF). Variables t and $\mathbf{x}$ are time and space coordinates, respectively. The velocity set used is the regular two-dimensional nine velocities scheme (D2Q9):

$$\mathbf{c}_i = \begin{cases} (0,0), & i = 0, \\ (c,0), (0,c), (-c,0), (0,-c), & i = 1, ..., 4, \\ (c,c), (-c,c), (-c,-c), (c,-c), & i = 5, ..., 8, \end{cases} \quad (2)$$

where $c$ is the lattice speed defined as $c = \Delta x / \Delta t$. The parameters $\Delta x$ and $\Delta t$ are the space and time steps.

The matrix $\mathbf{M}$ converts $(f_j - f_j^{eq})$ into a set of physical moments. In this study $\mathbf{M}$ assumes the same form used in Lallemand and Luo (2000). Then, the relaxation matrix becomes diagonal and can be written as

$$\mathbf{\Lambda} = \text{diag}\left( \tau_\rho^{-1}, \tau_e^{-1}, \tau_\varsigma^{-1}, \tau_j^{-1}, \tau_q^{-1}, \tau_j^{-1}, \tau_q^{-1}, \tau_\nu^{-1}, \tau_\nu^{-1} \right), \quad (3)$$

where the parameters $\tau$ are the different relaxation times for each physical moment.

The specific form of $f_i^{eq}$ can change depending on the particular method that is considered. For the single-phase weakly compressible LBM an usual form is (Krüger, Kusumaatmaja, Kuzmin, Shardt, Silva and Viggen, 2017):

$$f_i^{eq} = w_i \rho \left( 1 + \frac{c_{i\alpha}}{c_s^2} u_\alpha + \frac{c_{i\alpha} c_{i\beta} - c_s^2 \delta_{\alpha\beta}}{2 c_s^4} u_\alpha u_\beta, \right), \quad (4)$$

where terms $w_i$ are the weights related to each velocity $\mathbf{c}_i$ ($w_0 = 4/9$, $w_{1,2,3,4} = 1/9$ and $w_{5,6,7,8} = 1/36$) and $c_s$ is the fluid sound speed in the lattices, equal to $c/\sqrt{3}$ for the D2Q9 scheme. The variable $\rho$ is the density and $u_\alpha$ is the fluid velocity.

The term $F'_i$ on the right-hand side of Eq. (1) represents the forcing scheme. It adds the effects of an external force field, $F_\alpha$ to the macroscopic conservation equations. The specific form of forcing schemes used in this study will be detailed in the next subsections. For the single-phase LBM, usually the Guo, Zheng and Shi (2002) forcing scheme is adopted:

$$F'_i = w_i \left( \frac{c_{i\alpha}}{c_s^2} F_\alpha + \frac{c_{i\alpha}c_{i\beta} - c_s^2 \delta_{\alpha\beta}}{c_s^4} u_\alpha F_\beta \right). \tag{5}$$

The relation between particle distribution functions $f_i$ and the macroscopic fluid velocity $u_\alpha$ depends on the forcing scheme. For the schemes employed in this study, the density and velocity fields are given by

$$\rho = \sum_i f_i, \tag{6a}$$

$$\rho u_\alpha = \sum_i f_i c_{i\alpha} + \frac{F_\alpha \Delta t}{2}. \tag{6b}$$

The specific form of the source term $C_i$ depends on the particular model being used and will be detailed in the next subsections. Considering these equations with no source term ($C_i = 0$) and performing a second-order Chapman-Enskog analysis (Chapman and Cowling, 1990; Krüger et al., 2017), we obtain the mass and momentum conservation equations:

$$\partial_t \rho + \partial_\alpha(\rho u_\alpha) = 0, \tag{7a}$$

$$\partial_t(\rho u_\alpha) + \partial_\beta(\rho u_\alpha u_\beta) = -\partial_\alpha(\rho c_s^2) + \partial_\beta \sigma'_{\alpha\beta} + F_\alpha, \tag{7b}$$

where the viscous stress tensor $\sigma'_{\alpha\beta}$ can be written as

$$\sigma'_{\alpha\beta} = \rho \nu \left( \partial_\beta u_\alpha + \partial_\alpha u_\beta - \frac{2}{3} \delta_{\alpha\beta} \partial_\gamma u_\gamma \right) + \rho \nu_B \delta_{\alpha\beta} \partial_\gamma u_\gamma. \tag{8}$$

The kinematic and bulk viscosities $\nu$ and $\nu_B$, appearing in Eq. (8), are related to the relaxation times of LBM through

$$\nu = c_s^2 \left( \tau_\nu - \frac{\Delta t}{2} \right), \quad \nu_B = c_s^2 \left( \tau_e - \frac{\Delta t}{2} \right) - \frac{\nu}{3}. \tag{9}$$

Note that an isothermal ideal gas equation of state (EOS) $p_{EOS} = \rho c_s^2$ is recovered by the single-phase LBM.

## 2.1. Pseudopotential method

Shan and Chen (1993) proposed an interaction force based on nearest-neighbor interactions (see Shan (2008) for the definition of nearest-neighbor interactions):

$$F_\alpha^{SC} = -G\psi(\mathbf{x}) \sum_{i=1}^{8} \omega(|\mathbf{c}_i|^2) \psi(\mathbf{x} + \mathbf{c}_i \Delta t) c_{i\alpha} \Delta t, \tag{10}$$

where $\psi$ is a density-dependent interaction potential and G is a parameter that controls the strength of interaction. The parameters $\omega(|\mathbf{c}_i|^2)$ are weights, adopted as $\omega(1) = 1/3$ and $\omega(2) = 1/12$. The definition of interaction potential used in this study is the same as proposed by Yuan and Schaefer (2006), which allows the addition of arbitrary equations of state $p_{EOS}$ to the system:

$$\psi(\rho) = \sqrt{\frac{2(p_{EOS} - \rho c_s^2)}{Gc^2(\Delta t)^2}}. \tag{11}$$

When this technique is used, parameter G no longer controls the interaction strength. As authors usually adopt $p_{EOS} < \rho c_s^2$ in the literature, we can simply set $G = -1$.

The interaction force can be implemented by using several forcing schemes (Mapelli, Czelusniak, dos Santos Guzella and Cabezas-Gómez, 2022). One of the most used ones was proposed by Li et al. (2013):

$$\mathbf{MF'} = \begin{pmatrix} 0 \\ 6(u_x F_x + u_y F_y) + \frac{12\sigma^{\text{Li}} |\mathbf{F}_{\text{int}}|^2}{\psi^2(\tau_e - 0.5\Delta t)} \\ -6(u_x F_x + u_y F_y) - \frac{12\sigma^{\text{Li}} |\mathbf{F}_{\text{int}}|^2}{\psi^2(\tau_\varsigma - 0.5\Delta t)} \\ F_x \\ -F_x \\ F_y \\ -F_y \\ 2(u_x F_x - u_y F_y) \\ u_x F_y + u_y F_x \end{pmatrix}, \tag{12}$$

where $[\mathbf{F'}]_i = F'_i$. The term $\mathbf{F}_{\text{int}}$ is the interaction force, given by Shan and Chen (1993) force - Eq. (10). The total force is the combination of the interaction and gravitational forces, $\mathbf{F} = \mathbf{F}_{\text{int}} + \mathbf{F}_g$. The parameter $\sigma^{\text{Li}}$ is used to control the coexistence densities, for more details see Li et al. (2013).

Using the previous equations, it is possible to add an arbitrary equation of state into the method and control the coexistence densities. Moreover, it is also necessary to adjust the fluid surface tension. An approach was proposed by Li and Luo (2013), where source term assumes the form:

$$\mathbf{MC} = \begin{pmatrix} 0 \\ 1.5\tau_e^{-1}(Q_{xx} + Q_{yy}) \\ -1.5\tau_\varsigma^{-1}(Q_{xx} + Q_{yy}) \\ 0 \\ 0 \\ 0 \\ 0 \\ -\tau_\nu^{-1}(Q_{xx} - Q_{yy}) \\ -\tau_\nu^{-1}Q_{xy} \end{pmatrix}, \tag{13}$$

the variables $Q_{xx}$, $Q_{xy}$ and $Q_{yy}$ are calculated via

$$Q_{\alpha\beta} = \kappa^{\text{Li}} \frac{G}{2} \psi(\mathbf{x}) \sum_{i=1}^{8} \omega(|\mathbf{c}_i|^2) \big[ \psi(\mathbf{x} + \mathbf{c}_i \Delta t) - \psi(\mathbf{x}) \big] c_{i\alpha} c_{i\beta} (\Delta t)^2. \tag{14}$$

When analyzing the pseudopotential method numerical scheme using the third-order analysis proposed by Lycett-Brown and Luo (2015), we obtain the following mass and momentum conservation equations:

$$\partial_t \rho + \partial_\alpha (\rho u_\alpha) = 0, \tag{15a}$$

$$\partial_t (\rho u_\alpha) + \partial_\beta (\rho u_\alpha u_\beta) = -\partial_\beta p_{\alpha\beta} + \partial_\beta \sigma'_{\alpha\beta} + F_{g,\alpha}, \tag{15b}$$

where the viscous stress tensor $\sigma'_{\alpha\beta}$ is given by Eqs. (8) and (9). The interaction force effect was incorporated into the pressure tensor $p_{\alpha\beta}$, thus only the gravitational force $F_{g,\alpha}$ appears on Eq. (15b). The pressure tensor assumes the form:

$$p_{\alpha\beta} = \left( p_{EOS} + \frac{Gc^4(\Delta t)^4}{12} 24G\sigma^{\text{Li}}(\partial_\gamma \psi)(\partial_\gamma \psi) + \frac{Gc^4(\Delta t)^4}{12}(1 + 2\kappa^{\text{Li}})\psi \partial_\gamma \partial_\gamma \psi \right) \delta_{\alpha\beta} + \frac{Gc^4(\Delta t)^4}{6}(1 - \kappa^{\text{Li}})\psi \partial_\alpha \partial_\beta \psi. \tag{16}$$

From Eq. (16) we conclude that $\sigma^{\text{Li}}$ and $\kappa^{\text{Li}}$ are dimensionless variables.

For a planar interface problem, it is possible to derive an expression for the density profile (Shan and Chen, 1994; Shan, 2008; Krüger et al., 2017). A detailed derivation is presented at the Supplementary Material Appendix A. The P-LBM density profile is:

$$\left(\frac{d\rho}{dx}\right)^2 = \frac{8}{Gc^4(\Delta t)^4}\frac{\psi^\epsilon}{\dot{\psi}^2}\int_{\rho_v}^{\rho}(p_0 - p_{EOS})\frac{\dot{\psi}}{\psi^{1+\epsilon}}d\rho, \tag{17}$$

where $\epsilon = -16G\sigma^{Li}$ and $\dot{\psi} = d\psi/d\rho$. Another important relation is the P-LBM mechanical stability condition obtained from Eq. (17) (Li et al., 2013):

$$\int_{\rho_v}^{\rho_l}(p_{EOS} - p_0)\frac{\dot{\psi}}{\psi^{1+\epsilon}}d\rho = 0. \tag{18}$$

This relation implies that the value of $\epsilon = \epsilon(\sigma^{Li})$ can be adjusted to guarantee that the integral is satisfied for the same equilibrium densities $\rho_v$ and $\rho_l$ given by the Maxwell-rule (Callen, 1960; Bejan, 2016).

There are several variations of the pseudopotential method (P-LBM) in the lattice Boltzmann literature (Sbragaglia et al., 2007; Kupershtokh, Medvedev and Karpov, 2009; Czelusniak, Mapelli, Guzella, Cabezas-Gómez and Wagner, 2020). This work will be based on equations provided in this section, as they are among the most commonly used by authors in the literature.

## 2.2. Free energy method

There are several LBM schemes in the literature to implement the free energy method (FE-LBM) (Swift et al., 1996; Wagner, 2006; Siebert et al., 2014). The basic idea is to replace the divergence of the LBM original pressure tensor $-\partial_\alpha(\rho c_s^2)$ by a thermodynamic force of the form $F_\alpha = -\rho\partial_\alpha\mu$. The variable $\mu = \mu_b - \kappa\partial_\gamma\partial_\gamma\rho$ is the chemical potential with $\mu_b$ being the chemical potential in the bulk phase. The parameter $\kappa$ is used to control the surface tension. Then, when there is an unbalance in the chemical potential inside the system, a mass transfer process occur which tends to re-balance the different phases, towards equilibrium.

Most schemes proposed in the literature suffer from issues that prevent the equilibrium system from truly having a constant chemical potential throughout the domain. Recently Guo (2021) proposed a well-balanced free energy (WE-FE-LBM) scheme which is able to solve the issues from previous ones. These authors argued that the issues arise when a force like $F_\alpha = \partial_\alpha(\rho c_s^2)$ is added into the LBE to remove the original ideal gas EOS. Since both terms are space derivatives consisting in different discretizations, the remaining discretization errors contaminate the solution and prevent the system to attain a constant chemical potential.

The basic idea of the WB-FE-LBM is to avoid the appearance of the term $-\partial_\alpha(\rho c_s^2)$ at the macroscopic equations by modifying the equilibrium distribution function:

$$\begin{aligned} f_i^{eq} &= \rho - w_0\rho\frac{u_\gamma u_\gamma}{2c_s^2}, & i = 0, \\ f_i^{eq} &= w_i\rho\left[\frac{c_{i\alpha}}{c_s^2}u_\alpha + \frac{c_{i\alpha}c_{i\beta} - c_s^2\delta_{\alpha\beta}}{2c_s^2}u_\alpha u_\beta\right], & i \neq 0. \end{aligned} \tag{19}$$

The forcing scheme also must be modified to compensate the modifications in the equilibrium distribution function. The model was implemented using the BGK (Bhatnagar, Gross and Krook, 1954) collision operator. The WB-FE-LBM showed excellent results in terms of thermodynamic consistency (Guo, 2021). However, it suffered from numerical instability for low viscosity flows.

Zhang, Guo and Wang (2022a) proposed an improved well-balanced free energy method (IWB-FE-LBM) to increase the stability of the well-balanced method for low viscosity flows. The equilibrium distribution function is then written as:

$$\begin{aligned} f_i^{eq} &= \rho - (1-w_0)\frac{p_g}{c_s^2} - w_0\rho\frac{u_\gamma u_\gamma}{2c_s^2} - w_0\rho A\Delta t\partial_\gamma u_\gamma, & i = 0, \\ f_i^{eq} &= w_i\frac{p_g}{c_s^2} + w_i\rho\left[\frac{c_{i\alpha}}{c_s^2}u_\alpha + \frac{c_{i\alpha}c_{i\beta} - c_s^2\delta_{\alpha\beta}}{2c_s^4}(u_\alpha u_\beta + c_s^2 A\Delta t S_{\alpha\beta})\right], & i \neq 0, \end{aligned} \tag{20}$$

where $S_{\alpha\beta} = \partial_\alpha u_\beta + \partial_\beta u_\alpha$, $p_g = \rho_0 gy$, being $\rho_0$ a reference density used to model the buoyancy effect and $y$, the distance from a reference point in the y-direction. $A$ is a parameter to tune the viscosity.

The forcing scheme is modified to:

$$F_i = w_i \left[ \frac{c_{i\alpha}}{c_s^2} F_\alpha + \frac{(c_{i\alpha}c_{i\beta} - c_s^2 \delta_{\alpha\beta})}{c_s^4} u_\alpha (F_\beta - \partial_\beta p_g + c_s^2 \partial_\beta \rho) + \frac{1}{2}\left(\frac{|\mathbf{c}_i|^2}{c_s^2} - d\right)(u_\gamma \partial_\gamma \rho) \right], \tag{21}$$

where $\mathbf{F} = -\rho \partial_\alpha \mu + F_g$ is the total force and $\mathbf{F}_g = -(\rho - \rho_0) g \mathbf{j}$ denotes the buoyancy force. The model is implemented using the MRT collision operator. Through the Chapman-Enskog analysis, the resulting mass and momentum equations from the IWB-FE-LBM up to second order are:

$$\partial_t \rho + \partial_\gamma (\rho u_\gamma) = 0, \tag{22a}$$

$$\partial_t (\rho u_\alpha) + \partial_\beta (\rho u_\alpha u_\beta) = -\rho \partial_\alpha \mu + \partial_\beta \sigma'_{\alpha\beta} - F_{g,\alpha}. \tag{22b}$$

The viscous stress tensor is equal to Eq. (8). The viscosity coefficients are given by:

$$\nu = c_s^2 \left(\tau_\nu - \frac{\Delta t}{2} - A\right), \quad \nu_B = 2c_s^2 \left(\tau_e - \frac{\Delta t}{2} - \frac{A}{2}\right) - \frac{\nu}{3}. \tag{23}$$

Note that we are using the definition of bulk viscosity from Krüger et al. (2017) instead of Zhang et al. (2022a).

The discretization of the space derivatives are performed by using the 9-point finite difference stencils as follows:

$$\partial_\alpha \phi(\mathbf{x}) \approx \delta_\alpha \phi(\mathbf{x}) = \frac{1}{c_s^2 \Delta t} \sum_i \phi(\mathbf{x} + \mathbf{c}_i \Delta t) c_{i\alpha},$$
$$\partial_\gamma \partial_\gamma \phi(\mathbf{x}) \approx \delta_\gamma \delta_\gamma \psi(\mathbf{x}) = \frac{2}{c_s^2 \Delta t^2} \sum_i w_i \left[\phi(\mathbf{x} + \mathbf{c}_i \Delta t) - \phi(\mathbf{x})\right], \tag{24}$$

where the weights $w_i$ are the same as for the equilibrium distribution functions. Here, we use the symbol $\delta_\alpha$ to represent a numerical stencil which is an approximation of the real derivative $\partial_\alpha$.

Considering $\kappa$ as a constant, we can write the thermodynamic force as $F_\alpha = -\rho \partial_\alpha \mu = -\rho \partial_\alpha \mu_b + \kappa \rho \partial_\alpha \partial_\gamma \partial_\gamma \rho$. The bulk chemical potential can be related with the EOS by $-\rho \partial_\alpha \mu_b = -\partial_\alpha p_{EOS}$ and the entire pressure tensor is obtained from the relation $-\partial_\beta p_{\alpha\beta} = -\rho \partial_\alpha \mu$. Performing the necessary mathematical manipulations, the following pressure tensor is obtained:

$$p_{\alpha\beta} = \left(p_{EOS} - \frac{\kappa}{2}(\partial_\gamma \rho)(\partial_\gamma \rho) - \kappa \rho \partial_\gamma \partial_\gamma \rho\right)\delta_{\alpha\beta} + \kappa(\partial_\alpha \rho)(\partial_\beta \rho). \tag{25}$$

The EOS can be computed by $p_{EOS}(\rho, T) = (\rho \mu_b)|_0^\rho - \int_0^\rho \mu_b d\rho$.

Following the procedure presented in the Supplementary Material Appendix A, we obtain the following equation for the density space derivative in a planar interface problem:

$$\left(\frac{d\rho}{dx}\right)^2 = \frac{2\rho}{\kappa} \int_{\rho_v}^\rho (p_{EOS} - p_0) \frac{d\rho}{\rho^2}, \tag{26}$$

Using the boundary condition $d\rho/dx = 0$ for $\rho = \rho_l$:

$$\int_{\rho_v}^{\rho_l} (p_{EOS} - p_0) \frac{d\rho}{\rho^2} = 0. \tag{27}$$

This equation implies that the densities provided by the FE-LBM are equivalent to the ones given by the Maxwell-rule. Along the work, only the improved well-balanced free energy method will be employed and for simplicity, we will just call it by FE-LBM.

## 3. Methodology

The equation of state (EOS) adopted in this work is the Carnahan-Starling (C-S):

$$p_{EOS} = \rho RT \frac{1 + b\rho/4 + (b\rho/4)^2 - (b\rho/4)^3}{(1 - b\rho/4)^3} - a\rho^2, \tag{28}$$

where the parameters $a$, $b$ and $R$ are related to the critical properties: $a \approx 3.8533 p_c/\rho_c^2$, $b \approx 0.5218/\rho_c$ and $R = 2.7864 p_c/(\rho_c T_c)$. However, for our simulations we will simply adopt $a = 0.25$ or $a = 0.5$, $b = 4$ and $R = 1$ in lattice units (Baakeem, Bawazeer and Mohamad, 2021), since these are very common values employed in the literature (Czelusniak, Cabezas-Gómez and Wagner, 2023).

The values of $b$ and $R$ will no longer be mentioned in this work, as they are the same for all tests. When a parameter is given in lattice units, we will just use the generic representation "l.u.". For example, $a = 0.25$ l.u. or $a = 0.5$ l.u.

Considering the C-S EOS, its respective chemical potential is given by (Zhou and Huang, 2023):

$$\mu_b = RT \frac{3 - b\rho/4}{(1 - b\rho/4)^3} + RT \ln(\rho) + RT - 2a\rho. \tag{29}$$

For a fair comparison between methods, both P-LBM and FE-LBM should result in the same equilibrium densities, interface width, surface tension and sound speed. To obtain the same sound speed, we employ the EOS with the same $a$, $b$ and $R$ values for both methods. Next, to obtain the same phase densities, we have to set a $\sigma^{Li}$ value that guarantees the respect of the Maxwell-rule with the P-LBM. For performing this adjustment, one possibility is to run several simulations to set $\sigma^{Li}$. Instead, we prefer to compute this parameter theoretically using the procedure described in the Supplementary Material Appendix B. To match the surface tension, we specify a certain value of $\kappa$ in the FE-LBM and then we search for a $\kappa^{Li}$ that provides the same surface tension for the P-LBM. This procedure can also be done by running several simulations. Again, we prefer to compute this parameter theoretically following the Supplementary Material Appendix B. All the codes used in this work are publicly available, for more details check Supplementary Material.

## 4. Modified pseudopotential method

In the previous section we presented how to adjust the $\sigma^{Li}$ and $\kappa^{Li}$ parameters to match the equilibrium densities and surface tension between P-LBM and FE-LBM. However, it is still missing a procedure to match the interface width.

From Eq. (17), we see that the P-LBM density profile is a function of $p_{EOS}$, and $\epsilon$. However, $\epsilon$ is adjusted to match the Maxwell-rule for a given EOS from Eq. (18). Thus, $\epsilon$ is a function of $p_{EOS}$. This implies that the interface profile depends only on $p_{EOS}$. Since we are maintaining the same EOS parameters for the FE-LBM and P-LBM to achieve the same sound speed, changing the EOS parameters to control the P-LBM interface width is not an option.

To allow an adjustable interface width in the P-LBM without changing the EOS parameters, we introduce some modifications that are stated in next.

First we modify $Q_{\alpha\beta}$ of Eq. (14) to:

$$Q_{\alpha\beta} = \psi(\mathbf{x}) \sum_{i=1}^{8} \omega(|\mathbf{c}_i|^2) C_{i\alpha\beta} \left[ \psi(\mathbf{x} + \mathbf{c}_i \Delta t) - \psi(\mathbf{x}) \right], \tag{30a}$$

$$C_{i\alpha\beta} = C_1 G c^2 (\Delta t)^2 \delta_{\alpha\beta} + C_2 (c_{i\alpha} c_{i\beta} - c_s^2 \delta_{\alpha\beta})(\Delta t)^2 + \kappa^{Li} \frac{G}{2} c_{i\alpha} c_{i\beta} (\Delta t)^2, \tag{30b}$$

where the coefficients are given by:

$$C_1 = \frac{5}{6}(\kappa^P - 1), \tag{31a}$$

$$C_2 = \frac{1}{2}(1 - \kappa^P), \tag{31b}$$

We also modify the forcing scheme to:

$$\mathbf{MF}' = \begin{pmatrix} 0 \\ 6\left(u_x F_x + u_y F_y\right) + \frac{12\kappa^P \sigma^{\text{Li}}|\mathbf{F}_{\text{int}}|^2}{\psi^2 \Delta t(\tau_e - 0.5)} \\ -6\left(u_x F_x + u_y F_y\right) - \frac{12\kappa^P \sigma^{\text{Li}}|\mathbf{F}_{\text{int}}|^2}{\psi^2 \Delta t(\tau_\varsigma - 0.5)} \\ F_x \\ -F_x \\ F_y \\ -F_y \\ 2\left(u_x F_x - u_y F_y\right) \\ u_x F_y + u_y F_x \end{pmatrix}, \tag{32}$$

where the additional terms proposed by Li et al. (2013) were multiplied by $\kappa^P$. Then, the new pressure tensor is written as:

$$p_{\alpha\beta} = \left(p_{EOS} + \frac{Gc^4(\Delta t)^4}{12} 24 G \kappa^P \sigma^{\text{Li}} (\partial_\gamma \psi)(\partial_\gamma \psi) + \frac{Gc^4(\Delta t)^4}{12}(\kappa^P + 2\kappa^{\text{Li}})\psi \partial_\gamma \partial_\gamma \psi\right)\delta_{\alpha\beta} \\ + \frac{Gc^4(\Delta t)^4}{6}(\kappa^P - \kappa^{\text{Li}})\psi \partial_\alpha \partial_\beta \psi. \tag{33}$$

The $\epsilon$ value for the new pressure tensor remains $\epsilon = -16G\sigma^{\text{Li}}$, which is the same as before (check Supplementary Material Appendix C). This means that the equilibrium densities are not affected by the proposed modifications. The new parameter $\kappa^P$ is used to adjust the interface width while $\kappa^{\text{Li}}$ is used to adjust surface tension.

We call the new scheme as modified pseudopotential (MP-LBM). In the next example, the MP-LBM is tested in a planar interface problem with length $L_x = 100\Delta x$. The EOS defined in Eq. (28) is used with $a = 0.5$ l.u. The reduced temperature is 0.7. The $\sigma^{\text{Li}}$ parameter is computed using the procedure described in Supplementary Material Appendix B and $\kappa^{\text{Li}}$ is set to 0.

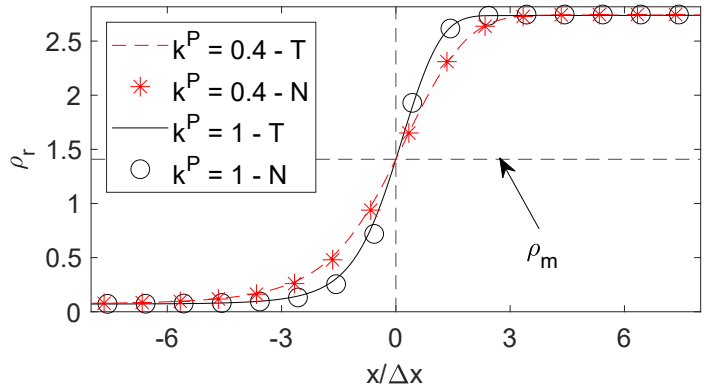

**Figure 1:** MP-LBM density profile with $\kappa^P = 0.4$ and $\kappa^P = 1$ for the C-S EOS with $a = 0.5$ l.u. and $T_r = 0.7$. Symbol T means theoretical results, N numerical and $\rho_m = (\rho_v + \rho_l)/2$.

Figure (1) shows the results considering two values of $\kappa^P$. First, we observe that the change in the $\kappa^P$ successfully changes the interface width without affect the equilibrium densities of each phase. Second, the LBM numerical results showed good agreement with the theoretical solution obtained from solving numerically Eq. (17). Since, the MP-LBM is the only pseudopotential scheme that will be used in the next examples, we will refer to that only as P-LBM.

## 5. Results
### 5.1. Static planar interface

The first test consists in simulating the equilibrium of two phases separated by a planar interface. After the system reaches the equilibrium, the densities obtained in the numerical simulations will be compared with the Maxwell-rule.

Although it was presented theoretical predictions for the equilibrium densities, Eq. (18) for the P-LBM and Eq. (27) for the FE-LBM, they do not guarantee that the numerical models will be able to full recover these values in the simulations. Since we are working with numerical methods, discretization errors can influence the method solution.

To better understand the method behaviour, we need theoretical predictions that will take into account the discretization errors. Those errors could be obtained from the LBE through a Chapman-Enskog analysis (Chapman and Cowling, 1990; Krüger et al., 2017) or by recursive substitutions (Lycett-Brown and Luo, 2015). However, these procedures are very difficult to be performed for higher orders. An easier approach consists in obtain a discrete equation in terms of the macroscopic properties (Peng et al., 2020; Guo, 2021). Then, this equation can be easily expanded using Taylor series until any desired order. We followed this approach and derived more precise expressions for the mechanical stability condition for the FE-LBM and P-LBM. The derivations are presented in the Supplementary Material Appendix D and E, respectively.

Following the derivation presented in the Supplementary Material Appendix D, we reach in the following equation for the FE-LBM planar interface (assuming $\tau_v = \tau_e$):

$$\int_{\rho_v}^{\rho_l} (p_{EOS} - p_0) \frac{d\rho}{\rho^2} = O(\Delta x^4). \tag{34}$$

Which means that Eq. (27) is an approximation of at least order $O(\Delta x^4)$. A similar theoretical analysis is performed with the P-LBM (also using the simplifying assumption $\tau_v = \tau_e$), the details are presented in the Supplemental Material Appendix E. From the analysis, it is concluded that:

$$\int_{\rho_v}^{\rho_l} (p_{EOS} - p_0) \frac{\dot{\bar{\psi}}}{\bar{\psi}^{1+\epsilon}} d\rho = O(\Delta x^2), \tag{35}$$

where $\bar{\psi} = \Delta x \psi$ was adopted to eliminate the influence of $\Delta x$ in the definition of $\psi$, Eq. (11). We observe that the discretization errors in the P-LBM mechanical stability condition are of second order while, for the FE-LBM - Eq. (34) - they are of fourth order. This fact potentially can make the errors in the P-LBM be higher than the FE-LBM.

To validate the theoretical analysis, numerical simulations of a planar interface were performed. A domain of $L_x = 100\Delta x$ was considered. In FE-LBM, we set $A = 0$ and all $\tau$ are equal to $\Delta t$, except $\tau_e$, which is equal to $0.75\Delta t$. These choices result in $\nu = c_s^2 \Delta t/2$ and $\nu_B = c_s^2 \Delta t/3$. For the P-LBM, we set $\tau_v = \Delta t$ and $\tau_e = \Delta t$, which provide the same values of $\nu$ and $\nu_B$ used in the FE tests . The value of $\kappa = 0.5$ l.u. was adopted in the FE-LBM and all other P-LBM parameters were chosen to match the FE simulation. The planar interface density profile was initialized using the same procedure as Czelusniak et al. (2020). Results for $a = 0.25$ l.u. and $a = 0.5$ l.u. in the C-S EOS are presented in Fig. (2) and compared with the Maxwell-rule.

As expected from the theoretical analysis, in equilibrium, the FE-LBM results - displayed in Fig. (2.a) - are in close agreement to the Maxwell rule. However, the lower stable temperature was $T_r = 0.65$ for $a = 0.5$ l.u. and $T_r = 0.6$ for $a = 0.25$ l.u., evidencing an stability issue. Zhou and Huang (2023) were able to achieve a temperature of 0.55 for the WB-FE-LBM with a different choice of parameters.

Regarding the P-LBM results displayed in Fig. (2.b) we see that the method is stable for $T_r = 0.5$ (lower $T_r$ were not tested) for both $a$ values in C-S EOS. For $T_R$ lower than 0.6, the method starts to loose accuracy due to discretization errors. For these lower temperatures, we conclude that we cannot use the $\sigma^{Li}$ values obtained theoretically from the mechanical stability condition. Unless we write a code to compute the theoretical solution considering a more accurate version of the mechanical stability condition (which we consider a cumbersome process), these $\sigma^{Li}$ values must be obtained empirically running previous LBM simulations.

Next, we evaluate the effect of varying $\tau_v$ in respect to $\tau_e$ and how it affects the equilibrium densities. For this analysis we defined the error of the vapor density in respect to the given by Maxwell rule $\epsilon_v = 100|\rho_v^N - \rho_v^M|/\rho_v^M$, where $\rho_v^N$ is the numerical vapor density obtained in simulation and $\rho_v^M$, the Maxwell rule vapor density. The results of this test are presented in Table (1). As we can realize, for the FE-LBM, the vapor error was not affected by the change of $\tau_v$ in respect to $\tau_e$.

For the pseudopotential method we observe a large variation of vapor density in respect to $\tau_v$. This problem was reported by Wu et al. (2018). The authors proposed a correction for this issue which consists in adopting $\tau_q = \Delta t/1.99$. We tested this approach and the results are in third line of Table 1. We conclude that this strategy really solves the problem.

Fundamental comparison between the pseudopotential and the free energy lattice Boltzmann methods

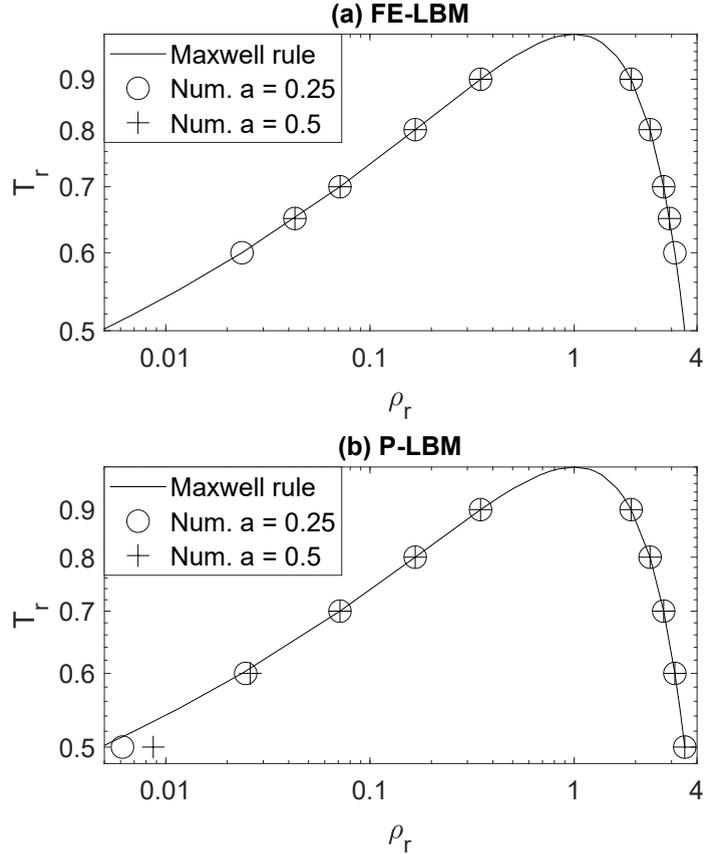

**Figure 2:** Comparison between numerical (Num.) results and the Maxwell-rule for the C-S EOS with $a = 0.25$ l.u. and $a = 0.5$ l.u. (a) FE-LBM with $\kappa = 0.5$ l.u. (b) P-LBM with parameters adjusted to match FE-LBM.

**Table 1**
Vapor density error in percentage for different $\tau_v^* = \tau_v/\Delta t$. In FE-LBM all other $\tau$ are equal to $\Delta t$ and $A = 0$. In P-LBM $\tau_q$ was also varied and the other $\tau$ are equal to $\Delta t$. Other simulation parameters are $T_r = 0.65$, $a = 0.5$ l.u. and $\kappa = 0.5$ l.u. The value of $\kappa^{Li}$ is equal to zero and $\kappa^P$ is adjusted to maintain the same interface width of FE-simulations.

| - | $\tau_v^* = 0.55$ | $\tau_v^* = 0.75$ | $\tau_v^* = 1$ | $\tau_v^* = 1.2$ |
|---|---|---|---|---|
| FE-LBM $\tau_q^* = 1$ | 0.0605 | 0.0605 | 0.0605 | 0.0605 |
| P-LBM $\tau_q^* = 1$ | 50.6 | 40.0 | 27.6 | 18.4 |
| P-LBM $\tau_q^* = 1/1.99$ | 27.7 | 27.7 | 27.6 | 27.6 |

### 5.2. Static Droplet

To evaluate the performance of the methods for curved interfaces, static droplet tests were performed. First, we compare the surface tension measured in the numerical tests (by applying the Young-Laplace law) with the theoretical one (described in Supplementary Material Appendix B). Then, the equilibrium densities of the static droplet tests are compared with the thermodynamic consistent results. These theoretical densities can be computed using the procedure described in Czelusniak et al. (2020), which is different from the Maxwell rule.

Numerical simulations of a 2D static droplet were performed. A domain of $L_x = L_y = 120\Delta x$ was used. For the FE-LBM we set $\kappa = 0.5$ l.u., $A = 0$ and all $\tau$ equal to $\Delta t$, except $\tau_e$ which is equal to $0.75\Delta t$. These choices result in $\nu = c_s^2 \Delta t/2$ and $\nu_B = c_s^2 \Delta t/3$. Also, $a = 0.25$ l.u. in the C-S EOS. The initialization procedure is similar to the one used by Czelusniak et al. (2020).

The P-LBM test was set similarly to the FE-LBM. Here we set all $\tau = \Delta t$ with the exception of $\tau_q = \Delta t/1.99$ to avoid the issues discussed in the previous subsection. The values of $\kappa^P$ and $\kappa^{Li}$ were adjusted to guarantee that the P-LBM has the same interface width and surface tension than for the FE-LBM simulations.

The pressure difference from inside to outside the droplet was measured and the results for different reduced temperatures and radiuses are shown in Fig. (3). The numerical results presented excellent agreement with the theoretical solution for both methods.

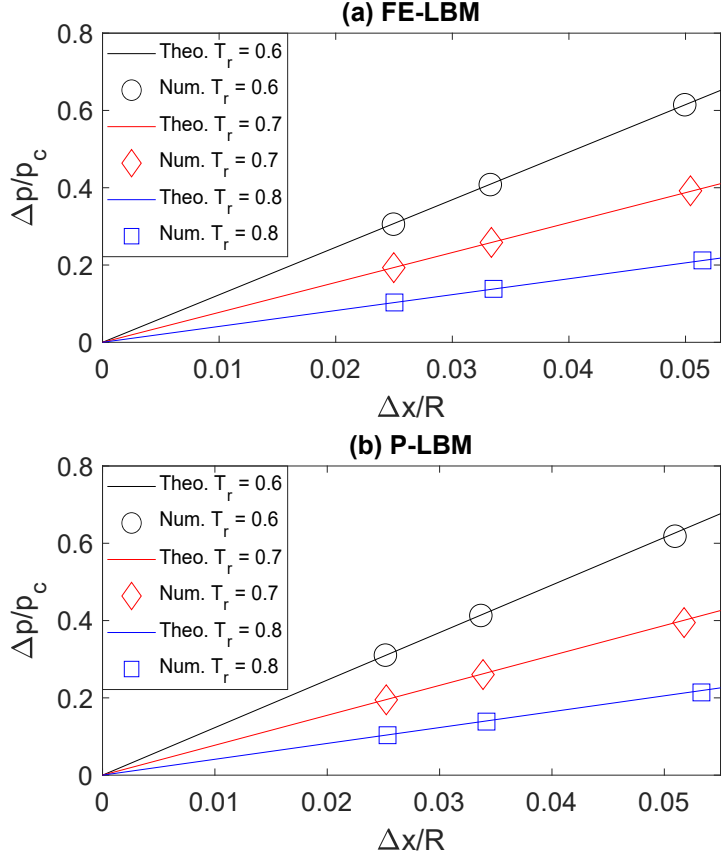

**Figure 3:** Young-Laplace test comparing numerical (Num.) and theoretical (Theo.) results for the C-S EOS with $a = 0.25$ l.u. (a) FE-LBM with $\kappa = 0.5$ l.u. (b) P-LBM with parameters selected to match the FE-LBM simulations.

Next, we compare the densities obtained in the numerical tests with the theoretical thermodynamic consistent densities. To expose the results in a concise way, we defined the reduced vapour density difference between the vapor phase density measured in the droplet test $\rho_v$ and the vapour density for the planar interface test $\rho_v^{PI}$ as $\Delta \rho_v = \rho_v - \rho_v^{PI}$ for the same temperature.

The case $\Delta x/R = 0$ correspond to a planar interface, then $\Delta \rho_v = 0$. When we increase $\Delta x/R$, the pressure difference between inside and outside the droplet starts to increase and the vapor density of the droplet test will deviate from the one obtained in the planar interface. In Fig. (4.a) we show the comparison between FE-LBM numerical and theoretical results. The agreement was excellent and the conclusion is that FE-LBM is thermodynamic consistent for the droplet test.

Regarding the P-LBM, in Fig. (3.b) the measured surface tension for $T_r = 0.6$ was close to the theoretical value. However, at this $T_r$, the planar interface test showed a discrepancy of 4% in respect to the Maxwell-rule (see Fig. (2.b)). Due to this difference, we decided to find the simulation parameters empirically rather than theoretically at this $T_r$. We found that $\epsilon = 1.9168$, $\kappa^{Li} = -1.70$ and $\kappa^P = 0.919$ provide an appropriate match between the planar interface results from P-LBM and from FE-LBM (considering $\kappa = 0.5$ l.u. in FE-LBM).

The droplet densities for the P-LBM are shown in Fig. (4.b). We can see that the P-LBM does not match perfectly with the consistent densities. However, the qualitative behaviour in which the vapor density increase with the curvature is reproduced. We also examined the absolute errors of the vapor density. For a droplet radius of $20\Delta x$ at $T_r = 0.8$, the thermodynamic consistent reduced vapor density is $\rho_{r;v} = \rho_v/\rho_c = 0.1824$, while the numerical density was $0.1786$, an error of approximately 2%.

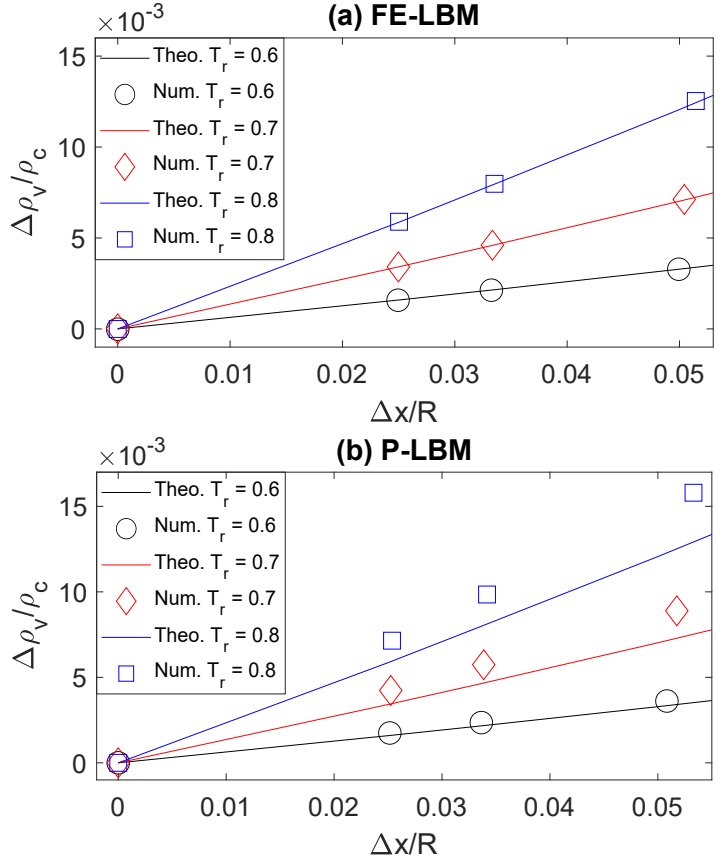

**Figure 4**: Droplet thermodynamic consistency test comparing numerical (Num.) and theoretica (Theo.) results for the C-S EOS with $a = 0.25$ l.u. (a) FE-LBM with $\kappa = 0.5$ l.u. (b) P-LBM with parameters adjusted to match FE-LBM simulation. $\Delta \rho_v = \rho_v - \rho_v^{\text{Pl}}$ is the vapor difference between the droplet and a planar interface.

This small error contradicts the results presented by Czelusniak, Mapelli, Wagner and Cabezas-Gómez (2022) that reported much larger errors for curved interface. However, until now we just analyzed the variation of the vapor density with the droplet radius. Czelusniak et al. (2020) found that the vapor densities in the P-LBM vary in a nonphysical way in respect to the surface tension. When the surface tension is reduced, the vapor density should approach the one obtained in a planar interface. However in the P-LBM the density deviates.

For a droplet at $T_r = 0.6$, we varied the surface tension by changing the $\kappa^{\text{Li}}$ parameter. The results are presented in Fig. (5). It is clear that the P-LBM was approaching the thermodynamic consistent results in Fig. (4.b) only for that specific choice of surface tension. For other surface tensions, the method largely deviates. For a droplet radius of $20\Delta x$ and surface tension $\gamma = 0.0046$ l.u. the numerical reduced vapor density is $0.0355$, while the theoretical one is $0.0251$, giving an error of 41%.

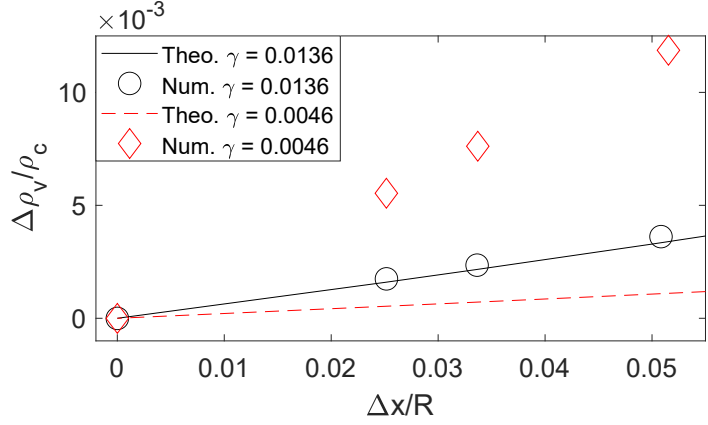

**Figure 5:** Droplet thermodynamic consistency test with P-LBM for the C-S EOS with $a = 0.25$ l.u., $T_r = 0.6$ and two different surface tensions. $\Delta \rho_v = \rho_v - \rho_v^{\text{Pl}}$ is the vapor difference between the droplet and a planar interface.

### 5.3. Two-phase flow between parallel plates

To evaluate both methods in a flow situation, a two-phase flow between parallel plates was selected. A domain of size $L_x = 101\Delta x$ and $L_y = 50\Delta x$ is used. The viscosity is $\nu = c_s^2 \Delta t/2$. The EOS parameters is $a = 0.25$ l.u. Initially half of the domain is filled with liquid (top) and the other half is filled with vapor (bottom). No gravity is considered here. A volumetric force in the horizontal direction is applied to drive the flow. The force $F_x$ magnitude depends on the simulation temperature: $1 \times 10^{-6}$ l.u., $5 \times 10^{-7}$ l.u. and $2 \times 10^{-7}$ l.u. for $T_r$ of 0.8, 0.7 and 0.6, respectively.

Periodic conditions were applied in the side boundaries and no-slip (Zou and He, 1997) were applied in top and bottom boundaries. The relaxation times for P-LBM are $\tau_\nu = \Delta t$, $\tau_e = \Delta t$ and $\tau_q = \Delta t/1.99$. To maintain the same kinematic and bulk viscosities in the FE-LBM we can set $A = 0$, $\tau_\nu = \Delta t$, $\tau_e = 0.75\Delta t$. However, we observed that the choice of $\tau_e = 0.75\Delta t$ in this problem can lead to an unstable simulation. Then, we set $\tau_e = \Delta t$ since the bulk viscosity is not important in this particular flow. We also set $\kappa = 0.5$ l.u. in FE-LBM and other P-LBM parameters were adjusted to match the same equilibrium densities, interface width and surface tension of the FE-LBM. The density profile is initialized in a similar way to the planar interface test. The initial distribution function is set as equal to the equilibrium one.

The LBM numerical results can be compared with two types of theoretical solutions. One solution considers a discrete interface and the jump conditions across this interface. However, the LBM is a diffuse interface method and will only approximate the discrete solution in the limit of a vanishing interface thickness. Then, we follow Zhang, Tang and Wu (2022b) and use the continuous theoretical solution which take into account the theoretical density profile. Both solutions are described in the Supplementary Material Appendix F.

The FE-LBM results are diplayed in Fig. (6). We observe that LBM indeed approximates better the continuous solution than the discrete one. In the previous tests, the FE-LBM was providing very close approximations to the theoretical solutions in the absence of flow. However, with flow we can observe the effect of numerical errors especially when the temperature is reduced to 0.6.

Before discuss the P-LBM results, recall that in Figs (4) and (5) we used $\sigma^{\text{Li}}$, $\kappa^P$ and $\kappa^{\text{Li}}$ obtained empirically rather than theoretically for $T_r = 0.6$. We did this to compensate the vapor density error in relation to the Maxwell rule. In this test, we will use the parameters obtained theoretically, since we want to do a direct comparison between the method and the theoretical solution.

The P-LBM results are displayed in Fig, (7). We observe that the deviations between the numerical and the theoretical solution are higher in comparison with the FE-LBM. These results suggest that LBM users should consider the necessity of applying a grid refinement procedure(Jaramillo, Mapelli and Cabezas-Gómez, 2022) to avoid large errors in flow simulations.

Finally, it was also tested a case with $T_r = 0.5$ and $F_x = 10^{-7}$ l.u. Only the pseudopotential method resulted in a stable simulation. Due to the lack of comparison between both methods, this case is not presented here.

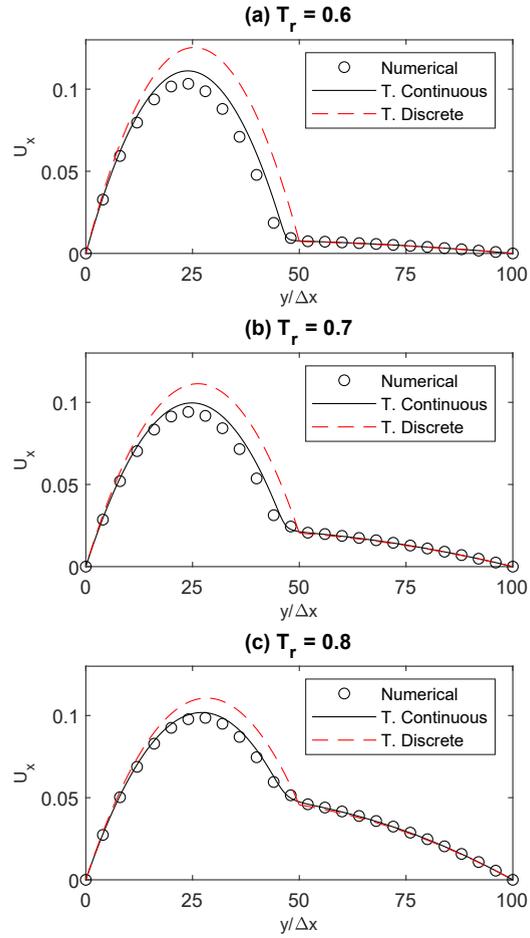

**Figure 6:** Two-phase flow between parallel plates for FE-LBM ($\kappa = 0.5$ l.u.) with the C-S EOS using $a = 0.25$ l.u.

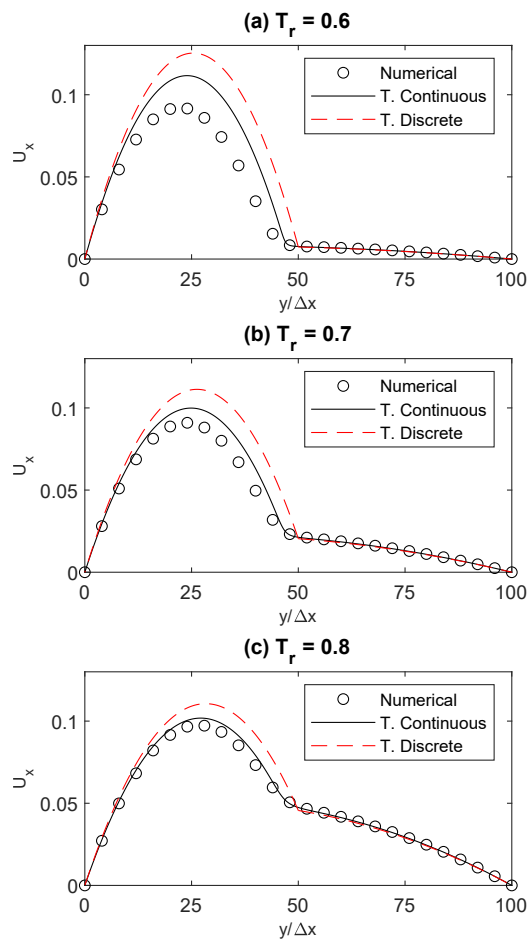

**Figure 7:** Two-phase flow between parallel plates for P-LBM with the C-S EOS using $a = 0.25$ l.u. The P-LBM parameters were set to match the FE-LBM simulation physical properties.

Fundamental comparison between the pseudopotential and the free energy lattice Boltzmann methods

## 5.4. Droplet Collision

A dynamic test was select to finish the study: the coalescence of two droplets. We considered a domain of $L_x = 200\Delta x$ and $L_y = 100\Delta x$. The viscosities are $\nu = c_s^2 \Delta t/2$ and $\nu_B = c_s^2 \Delta t/3$. The EOS parameter is $a = 0.25$ l.u. and the simulation occurs at $T_r = 0.8$. Initially two droplets of diameter $R_d = 50\Delta x$ are placed at positions $(50\Delta x, 50\Delta x)$ and $(150\Delta x, 50\Delta x)$.

A force acting in the x-axis of the form $F_x = a_x|\rho - \rho_{ave}|$ was used to move the droplets towards coalescence, where $\rho_{ave}$ is the average density in the domain. We set the acceleration $a_x = 5 \times 10^{-6} \Delta x (\Delta t)^{-2}$ for $x \leq 100\Delta x$ and $a_x = -5 \times 10^{-6} \Delta x (\Delta t)^{-2}$ for $x > 100\Delta x$. Then, the forces in each side of the domain are in opposite direction and make the droplets approach each other. The force is only applied after a time $t > 2000\Delta t$.

The parameters of FE-LBM are $A = 0$, $\tau_\nu = \Delta t$, $\tau_e = 0.75\Delta t$ and $\kappa = 0.5$. For the P-LBM $\tau_\nu = \Delta t$, $\tau_e = \Delta t$ and $\tau_q = \Delta t/1.99$. Other P-LBM parameters were adjusted to match the same equilibrium densities, interface width and surface tension of the FE-LBM. The droplets density profiles were initialized similarly to the static droplet case. The initial distribution function is considered to be equal to the equilibrium one.

In Fig. (8) we plotted the contours for $\rho_m = (\rho_v + \rho_l)/2$ (average density between liquid and vapor) at different dimensionless times $t^* = t\sqrt{a_d/D_d}$. First, we observe that both methods provided different results. At $t^* = 1.923$ coalescence was almost beginning for both methods. At $t^* = 1.935$ the droplets in P-LBM already merged into a single droplet, while in the FE-LBM they did not. Only at $t^* = 1.961$ the FE-LBM droplets merged.

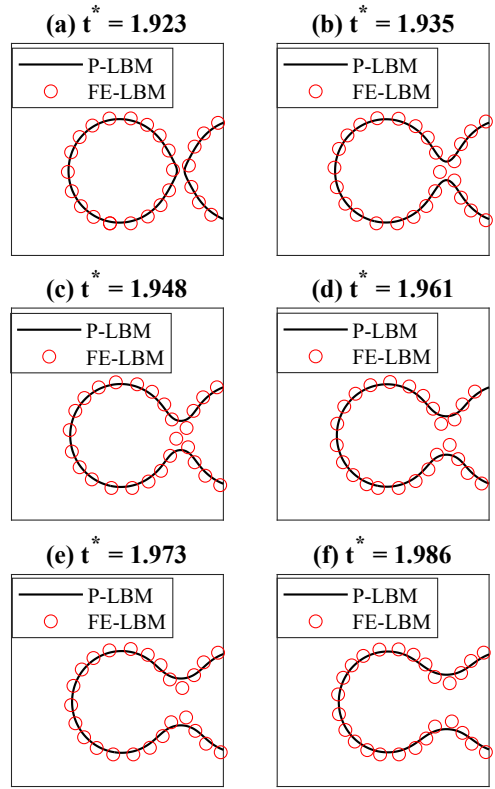

**Figure 8:** Droplet collision test with P-LBM and FE-LBM for the C-S EOS with $a = 0.25$ l.u. and $T_r = 0.8$.

Next, we decided to compare both simulations but now adding a time delay between them. The P-LBM snapshots are shown at the same time $t^*$ as in the previous image, while the snapshots of FE-LBM are of time $t^* + 0.0253$. This delay was carefully chosen to make the simulation of both methods merge the droplets together. Then, we can compare the dynamics after merging. These results are presented in Fig. (9).

We observe that both methods provided the same droplet dynamics after merging. These results suggest that the differences between methods only happens before the merging process is complete. More studies are needed to better describe this phenomenon; however, some thoughts can be raised about this.

First, we are initializing the droplets at a certain distance and due to the action of the force they approach each other. In principle the droplets interfaces should get in contact at the same time for both methods. But due to high order discretization errors in both methods, the contact time can differ by a small amount. Since the merging process is very fast as droplets get in contact, if one method starts the merging slightly before, this can produce a visible difference in the method results. So, one possible explanation can be related with the droplets starting the merging at small different times.

Other explanation can be associated with the merging process itself. When two interfaces are merging, the dynamics of this process will be ruled by the pressure tensor of each method. Since both pressure tensors are different, the speed of process can be different for each method.

After that, the dynamics of the simulation will be ruled by properties as surface tension, viscosity and density which are the same for both methods. This explain why results are very similar when we delay simulations to synchronize the merging process.

Finally, we repeat the simulation for other temperatures. We maintain the same $\kappa = 0.5$ l.u. for the FE-LBM and adjust the P-LBM parameters to match physical properties with FE method. For $T_r = 0.7$, the FE-LBM simulation become stable at a certain moment in the merging process. With the P-LBM we were able to produce a stable simulation with $T_r = 0.6$. We also tested $T_r = 0.5$ which made the P-LBM unstable. This test also show the superior stability of P-LBM over the well balanced FE-LBM.

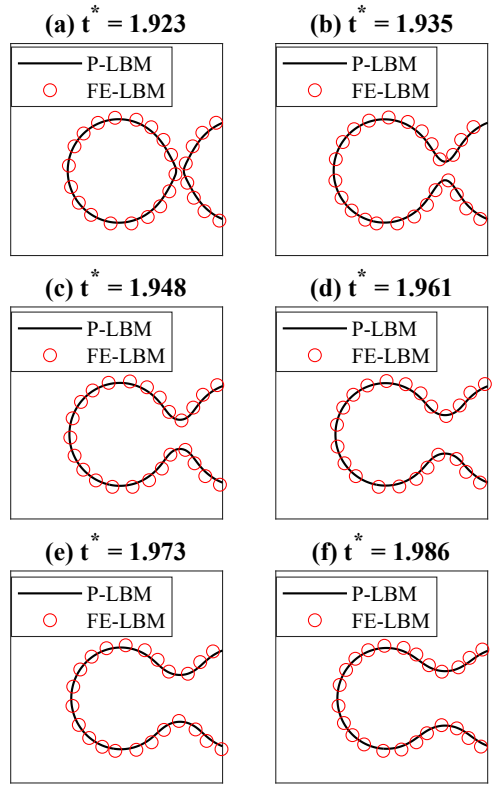

**Figure 9:** Droplet colision test with P-LBM and FE-LBM for the C-S EOS with $a = 0.25$ l.u. and $T_r = 0.8$. FE-LBM snapshots are delayed by $\Delta t^* = 0.0253$.

## 6. Conclusion

In this paper we performed a comparison between two methods for multiphase single-component fluid simulations: the pseudopotential (P) and the free energy (FE) lattice Boltzmann methods (LBM). For the FE method we selected the improved well balanced scheme. The equation of state (EOS) considered for this study is the Carnahan-Starling (C-S). We also presented a novel approach to control the interface width of the P-LBM without changing the EOS parameters.

From the planar interface tests, we noticed that the FE-LBM automatically results in the Maxwell rule, while for the P-LBM it is necessary to adjust the $\sigma^{\text{Li}}$ parameter to obtain the correct densities. For a certain range of temperatures it is possible to obtain this parameter theoretically, avoiding a costly empirical trial and error process. However for reduced temperatures of $T_r = 0.6$ or lower, the theoretical solution do not work well, and an empirical process is necessary any way. However, differently of the FE-LBM, the P-LBM allowed the simulation with lower reduced temperatures, e.g. $T_r = 0.6$ or bellow considering the choice of parameters in our study.

In droplet tests we observed that the FE-LBM maintains the thermodynamic consistency, while the P-LBM fails. The adjustment carried out with the $\sigma^{\text{Li}}$ parameter in the planar interface test is not enough to guarantee the correct densities for droplets, specially when the surface tension is varied by adjusting the $\kappa^{\text{Li}}$ parameter.

A two-phase flow between parallel plates was simulated to compare both methods in a flow situation. We observed that the FE-LBM do not reproduced the same accuracy showed in the previous tests without flow. However its accuracy was superior compared with the P-LBM. The drop in accuracy for low temperatures indicates that mesh refinement procedures must be considered in simulations of real physical systems.

In the droplet collision simulation, we observed that after the merging of two droplets, the interface dynamics is very similar for both methods. However, it was noticed a difference between methods before merging is complete. Due to this effect, the coalescence process occurs in less time for the P-LBM.

We conclude that the FE-LBM is more practicable and accurate than the P-LBM in the temperature range where it converged. However, for applications that demand lower reduced temperatures, the FE-LBM may not be suitable. In this situation, P-LBM can be more feasible.

It should be considered that the present comparison was only possible due to the proposed modified P-LBM, that allows to control the interface width while maintaining the EOS parameters. This approach made possible the results comparison on the same base. The novel approach is also promising for simulating physical problems employing other EOS under a controlled interface width. This can be important for obtaining numerical solutions of multiphase phase change problems.

## Supplementary Material

This paper is accompanied by supplementary material containing 6 appendices mentioned in the text. Appendix A shows the derivation planar interface solutions; Appendix B shows how to compute simulation parameters theoretically; Appendix C shows detailed derivations related to the modified pseudopotential method; Appendix D and E show derivations of planar interface relations with error terms and Appendix F shows solutions for two-phase flow between parallel plates.

## Acknowledgments

We gratefully acknowledge the support of the ALFA - Artificial Lift and Flow Assurance Research Group, hosted by the Center for Energy and Petroleum Studies (CEPETRO) at the Universidade Estadual de Campinas (UNICAMP), Brazil. We also thank EPIC – Energy Production Innovation Center, hosted by the Universidade Estadual de Campinas (UNICAMP) and sponsored by Equinor Brazil and FAPESP – São Paulo Research Foundation (2017/15736-3, 2022/08305-4 and 2023/02383-6) for their funding and support, as well as the ANP (Brazil's National Oil, Natural Gas, and Biofuels Agency) for assistance through the R&D levy regulation. Special thanks to the CEPETRO-UNICAMP and FEM-UNICAMP for their collaboration, and to São Carlos School of Engineering - University of São Paulo (EESC-USP) for their contributions.

## AUTHOR DECLARATIONS
### Conflict of Interest
The authors have no conflicts to disclose.

### Data Availability Statement
The data that support the findings of this study are available from the corresponding author upon reasonable request. All the codes used in this work are publicly available at https://github.com/luizeducze/SuppMaterial_Alfa01.

# CRediT authorship contribution statement

**Luiz Eduardo Czelusniak:** Conceptualization (equal), Data curation (equal), Formal analysis (equal), Investigation (equal), Methodology (equal), Software (equal), Validation (equal), Visualization (equal), Writing - original draft (equal). **Ivan Talão Martins:** Data curation (equal), Formal analysis (equal), Investigation (equal), Methodology (equal), Software (equal), Validation (equal), Writing - review & editing (equal). **Luben Cabezas Gómez:** Conceptualization (equal), Formal analysis (equal), Methodology (equal), Supervision (equal), Validation (equal), Writing - review & editing (equal). **Natan Augusto Vieira Bulgarelli:** Conceptualization (equal), Methodology (equal), Project administration (equal), Resources (equal), Supervision (equal), Validation (equal), Writing - review & editing (equal). **William Monte Verde:** Conceptualization (equal), Methodology (equal), Project administration (equal), Resources (equal), Supervision (equal), Validation (equal), Writing - review & editing (equal). **Marcelo Souza de Castro:** Conceptualization (equal), Funding acquisition (equal), Methodology (equal), Project administration (equal), Resources (equal), Supervision (equal), Validation (equal), Writing - review & editing (equal).

# References

Baakeem, S.S., Bawazeer, S.A., Mohamad, A.A., 2021. A novel approach of unit conversion in the lattice boltzmann method. Applied Sciences 11, 6386.

Bejan, A., 2016. Advanced engineering thermodynamics. John Wiley & Sons.

Bhatnagar, P.L., Gross, E.P., Krook, M., 1954. A model for collision processes in gases. i. small amplitude processes in charged and neutral one-component systems. Physical review 94, 511.

Callen, H.B., 1960. Thermodynamics and an introduction to thermostatistics. John Wiley & Sons, New York .

Chapman, S., Cowling, T.G., 1990. The mathematical theory of non-uniform gases: an account of the kinetic theory of viscosity, thermal conduction and diffusion in gases. Cambridge university press.

Czelusniak, L.E., Cabezas-Gómez, L., Wagner, A.J., 2023. Effect of gravity on phase transition for liquid–gas simulations. Physics of Fluids 35, 043324.

Czelusniak, L.E., Mapelli, V.P., Guzella, M., Cabezas-Gómez, L., Wagner, A.J., 2020. Force approach for the pseudopotential lattice boltzmann method. Physical Review E 102, 033307.

Czelusniak, L.E., Mapelli, V.P., Wagner, A.J., Cabezas-Gómez, L., 2022. Shaping the equation of state to improve numerical accuracy and stability of the pseudopotential lattice boltzmann method. Physical Review E 105, 015303.

Fei, L., Yang, J., Chen, Y., Mo, H., Luo, K.H., 2020. Mesoscopic simulation of three-dimensional pool boiling based on a phase-change cascaded lattice boltzmann method. Physics of Fluids 32.

Guo, Z., 2021. Well-balanced lattice boltzmann model for two-phase systems. Physics of Fluids 33, 031709.

Guo, Z., Zheng, C., Shi, B., 2002. Discrete lattice effects on the forcing term in the lattice boltzmann method. Physical review E 65, 046308.

Huang, R., Wu, H., 2016. Third-order analysis of pseudopotential lattice boltzmann model for multiphase flow. Journal of Computational Physics 327, 121–139.

Jaramillo, A., Mapelli, V.P., Cabezas-Gómez, L., 2022. Pseudopotential lattice boltzmann method for boiling heat transfer: A mesh refinement procedure. Applied Thermal Engineering 213, 118705.

Kikkinides, E., Yiotis, A., Kainourgiakis, M., Stubos, A., 2008. Thermodynamic consistency of liquid-gas lattice boltzmann methods: Interfacial property issues. Physical Review E 78, 036702.

Krüger, T., Kusumaatmaja, H., Kuzmin, A., Shardt, O., Silva, G., Viggen, E.M., 2017. The lattice boltzmann method. Springer International Publishing 10, 4–15.

Kupershtokh, A.L., Medvedev, D., Karpov, D., 2009. On equations of state in a lattice boltzmann method. Computers & Mathematics with Applications 58, 965–974.

Lallemand, P., Luo, L.S., 2000. Theory of the lattice boltzmann method: Dispersion, dissipation, isotropy, galilean invariance, and stability. Physical review E 61, 6546.

Li, Q., Kang, Q., Francois, M.M., He, Y., Luo, K., 2015. Lattice boltzmann modeling of boiling heat transfer: The boiling curve and the effects of wettability. International Journal of Heat and Mass Transfer 85, 787–796.

Li, Q., Luo, K., 2013. Achieving tunable surface tension in the pseudopotential lattice boltzmann modeling of multiphase flows. Physical Review E 88, 053307.

Li, Q., Luo, K., Li, X., 2013. Lattice boltzmann modeling of multiphase flows at large density ratio with an improved pseudopotential model. Physical Review E 87, 053301.

Li, W., Li, Q., Yu, Y., Luo, K.H., 2021. Nucleate boiling enhancement by structured surfaces with distributed wettability-modified regions: A lattice boltzmann study. Applied Thermal Engineering 194, 117130.

Lycett-Brown, D., Luo, K.H., 2015. Improved forcing scheme in pseudopotential lattice boltzmann methods for multiphase flow at arbitrarily high density ratios. Physical Review E 91, 023305.

Mapelli, V.P., Czelusniak, L.E., dos Santos Guzella, M., Cabezas-Gómez, L., 2022. On the force scheme influence on pseudopotential method coexistence curve. Physica A: statistical mechanics and its applications 599, 127411.

Peng, C., Ayala, L.F., Ayala, O.M., Wang, L.P., 2019. Isotropy and spurious currents in pseudo-potential multiphase lattice boltzmann models. Computers & Fluids 191, 104257.

Peng, C., Ayala, L.F., Wang, Z., Ayala, O.M., 2020. Attainment of rigorous thermodynamic consistency and surface tension in single-component pseudopotential lattice boltzmann models via a customized equation of state. Physical Review E 101, 063309.

Sbragaglia, M., Benzi, R., Biferale, L., Succi, S., Sugiyama, K., Toschi, F., 2007. Generalized lattice boltzmann method with multirange pseudopotential. Physical Review E 75, 026702.

Shan, X., 2006. Analysis and reduction of the spurious current in a class of multiphase lattice boltzmann models. Physical Review E 73, 047701.

Shan, X., 2008. Pressure tensor calculation in a class of nonideal gas lattice boltzmann models. Physical Review E 77, 066702.

Shan, X., Chen, H., 1993. Lattice boltzmann model for simulating flows with multiple phases and components. Physical Review E 47, 1815.

Shan, X., Chen, H., 1994. Simulation of nonideal gases and liquid-gas phase transitions by the lattice boltzmann equation. Physical Review E 49, 2941.

Siebert, D., Philippi, P., Mattila, K., 2014. Consistent lattice boltzmann equations for phase transitions. Physical Review E 90, 053310.

Sudhakar, T., Das, A.K., 2020. Evolution of multiphase lattice boltzmann method: A review. Journal of The Institution of Engineers (India): Series C 101, 711–719.

Swift, M.R., Orlandini, E., Osborn, W., Yeomans, J., 1996. Lattice boltzmann simulations of liquid-gas and binary fluid systems. Physical Review E 54, 5041.

Wagner, A., 2006. Thermodynamic consistency of liquid-gas lattice boltzmann simulations. Physical Review E 74, 056703.

Wang, H., Lou, Q., Li, L., 2020. Mesoscale simulations of saturated flow boiling heat transfer in a horizontal microchannel. Numerical Heat Transfer, Part A: Applications 78, 107–124.

Wang, J., Liang, G., Yin, X., Shen, S., 2023. Pool boiling on micro-structured surface with lattice boltzmann method. International Journal of Thermal Sciences 187, 108170.

Wu, Y., Gui, N., Yang, X., Tu, J., Jiang, S., 2018. Fourth-order analysis of force terms in multiphase pseudopotential lattice boltzmann model. Computers & Mathematics with Applications 76, 1699–1712.

Yuan, P., Schaefer, L., 2006. Equations of state in a lattice boltzmann model. Physics of Fluids 18, 042101.

Zhang, C., Chen, L., Ji, W., Liu, Y., Liu, L., Tao, W.Q., 2021. Lattice boltzmann mesoscopic modeling of flow boiling heat transfer processes in a microchannel. Applied Thermal Engineering 197, 117369.

Zhang, C., Guo, Z., Wang, L.P., 2022a. Improved well-balanced free-energy lattice boltzmann model for two-phase flow with high reynolds number and large viscosity ratio. Physics of Fluids 34, 012110.

Zhang, S., Tang, J., Wu, H., 2022b. Phase-field lattice boltzmann model for two-phase flows with large density ratio. Physical Review E 105, 015304.

Zhou, Z.T., Huang, J.J., 2023. Study of single-component two-phase free energy lattice boltzmann models using various equations of state. Physics of Fluids 35.

Zou, Q., He, X., 1997. On pressure and velocity boundary conditions for the lattice boltzmann bgk model. Physics of fluids 9, 1591–1598.

# Supplementary Material: Theoretical and numerical comparison between the pseudopotential and the free energy lattice Boltzmann methods

Luiz Eduardo Czelusniak[a,*], Ivan Talão Martins[b], Luben Cabezas Gómez[b], Natan Augusto Vieira Bulgarelli[a], William Monte Verde[a] and Marcelo Souza de Castro[a]

[a]*Center for Energy and Petroleum Studies, State University of Campinas, Campinas, 13083-896, São Paulo, Brazil*
[b]*Department of Mechanical Engineering, São Carlos School of Engineering, University of São Paulo, São Carlos, 13566-590, São Paulo, Brazil*

ABSTRACT

In this supplementary material we show some lengthy mathematical derivations that complement the main paper and also the codes used in the numerical simulations. All the codes used in this work are publicly available at https://github.com/luizeducze/SuppMaterial_Alfa01. This link also contains a description of all the codes in the README.md file.

## A. Planar interface solution

In this section, density profile and equilibrium densities equations will be derived for planar interface cases following the procedures used by Shan and Chen (1994), Shan (2008) and Krüger, Kusumaatmaja, Kuzmin, Shardt, Silva and Viggen (2017). We use this theoretical solutions to compute simulation parameters in our study. Due to their importante we decided to present their derivation although it can also be found in other works. The starting point are the pressure tensors of the pseudopotential (P-LBM) and free energy (FE-LBM) methods. For the P-LBM numerical scheme provided by the works of Li and Luo (2013) and Li and Luo (2013) the pressure tensor obtained in a third order analysis (Lycett-Brown and Luo, 2015) is given by:

$$p_{\alpha\beta} = \left( p_{EOS} + \frac{Gc^4(\Delta t)^4}{12} 24G\sigma^{\text{Li}}(\partial_\gamma \psi)(\partial_\gamma \psi) + \frac{Gc^4(\Delta t)^4}{12}(1 + 2\kappa^{\text{Li}})\psi \partial_\gamma \partial_\gamma \psi \right)\delta_{\alpha\beta} \\ + \frac{Gc^4(\Delta t)^4}{6}(1 - \kappa^{\text{Li}})\psi \partial_\alpha \partial_\beta \psi. \qquad (1)$$

This equation is the same as Eq. (16) in the main paper.

Now, it will be considered a situation where two phases separated by a planar interface (along the x-direction) are in equilibrium. In this case, the pressure component $p_{xx}$ is constant along the interface and we define $p_{xx} = p_0$. Equation 1 can be rewritten considering that there are density gradients only in the x-direction:

$$p_0 = p_{EOS} + \frac{Gc^4(\Delta t)^4}{12}\left[ A\left(\frac{d\psi}{dx}\right)^2 + B\psi \frac{d^2\psi}{dx^2} \right], \qquad (2)$$

where the parameters $A$, and $B$ are obtained from the coefficients of Eq. (1): $A = 24G\sigma^{\text{Li}}$, $B = 3$.

Assuming that $\psi$ is a monotonic growing function in respect to $x$, then for each value of $x$ there is only one value of $\psi$ and vice-versa. Then, $d\psi/dx$ can be written as a function of $\psi$: $d\psi/dx = z(\psi)$. In this way:

$$\frac{d^2\psi}{dx^2} = \frac{d}{dx}z(\psi) = \frac{dz}{d\psi}\frac{d\psi}{dx} = \dot z z = \frac{1}{2}\frac{dz^2}{d\psi} \qquad (3)$$

---

*Corresponding author

✉ luizedcz@unicamp.br (L.E. Czelusniak); ivanmartins@usp.br (I.T. Martins); lubencg@sc.usp.br (L.C. Gómez); (N.A.V. Bulgarelli); (W.M. Verde); (M.S.d. Castro)

ORCID(s): 0000-0001-8488-5720 (L.E. Czelusniak); 0000-0003-3961-5638 (I.T. Martins); 0000-0002-9550-9453 (L.C. Gómez); 0000-0001-7565-3565 (N.A.V. Bulgarelli); 0000-0003-0353-3868 (W.M. Verde); 0000-0001-9797-9144 (M.S.d. Castro)

Now, Eq. (2) can be rewritten as:

$$\frac{dz^2}{d\psi} - \epsilon \frac{z^2}{\psi} = \psi^\epsilon \frac{d}{d\psi}\left(\frac{z^2}{\psi^\epsilon}\right) = \frac{12}{Gc^4(\Delta t)^4}\frac{2(p_0 - p_{EOS})}{B\psi}, \tag{4}$$

where $\epsilon = -2A/B$. Solving this equation and applying the boundary condition $z(\rho_v) = 0$, we obtain:

$$\left(\frac{d\psi}{dx}\right)^2 = (\dot\psi)^2\left(\frac{d\rho}{dx}\right)^2 = \frac{12\psi^\epsilon}{Gc^4(\Delta t)^4}\int_{\psi_v}^{\psi}\frac{2(p_0 - p_{EOS})}{B}\frac{d\psi}{\psi^{1+\epsilon}} = \frac{12\psi^\epsilon}{Gc^4(\Delta t)^4}\int_{\rho_v}^{\rho}\frac{2(p_0 - p_{EOS})}{B}\frac{\dot\psi}{\psi^{1+\epsilon}}d\rho, \tag{5}$$

where $\dot\psi = d\psi/d\rho$. The conversion of the integral in respect to $\psi$ into an integral in respect to $\rho$ is only possible if $\rho$ is also a monotonic growing function in respect to $x$. Also, from the boundary condition $z(\rho_l) = 0$, we obtain:

$$\int_{\rho_v}^{\rho_l}(p_{EOS} - p_0)\frac{\dot\psi}{\psi^{1+\epsilon}}d\rho = 0. \tag{6}$$

This equation is known as the mechanical stability condition for the P-LBM.

For the FE-LBM, the pressure tensor is given by the following equation:

$$p_{\alpha\beta} = \left(p_{EOS} - \frac{\kappa}{2}(\partial_\gamma\rho)(\partial_\gamma\rho) - \kappa\rho\partial_\gamma\partial_\gamma\rho\right)\delta_{\alpha\beta} + \kappa(\partial_\alpha\rho)(\partial_\beta\rho). \tag{7}$$

This equation is equivalent to Eq. (25) of the main paper. Following the same procedure used for the P-LBM and considering a planar interface along the x-direction with a constant pressure component $p_{xx} = p_0$, we can rewritte this equation to:

$$p_0 = p_{EOS} + \kappa\left[\frac{1}{2}\left(\frac{d\rho}{dx}\right)^2 - \rho\frac{d^2\rho}{dx^2}\right], \tag{8}$$

Using the same strategy as before, we define a function $z(\rho) = d\rho/dx$. This is valid if $\rho$ is a monotonic function in respect to $x$. Then, we obtain the following relation:

$$\frac{d^2\rho}{dx^2} = \frac{d}{dx}z(\rho) = \frac{dz}{d\rho}\frac{d\rho}{dx} = \dot z z = \frac{1}{2}\frac{dz^2}{d\rho} \tag{9}$$

Replacing this result into Eq. (8):

$$\frac{z^2}{\rho} - \frac{dz^2}{d\rho} = -\rho\frac{d}{d\rho}\left(\frac{z^2}{\rho}\right) = \frac{2(p_0 - p_{EOS})}{\kappa\rho}, \tag{10}$$

Performing the integration and applying the boundary condition $z(\rho_v) = 0$, we obtain:

$$\left(\frac{d\rho}{dx}\right)^2 = \frac{2\rho}{\kappa}\int_{\rho_v}^{\rho}(p_{EOS} - p_0)\frac{d\rho}{\rho^2}, \tag{11}$$

Using the boundary condition $d\rho/dx = 0$ for $\rho = \rho_l$:

$$\int_{\rho_v}^{\rho_l}(p_{EOS} - p_0)\frac{d\rho}{\rho^2} = 0. \tag{12}$$

## B. How compute $\sigma^{\text{Li}}$ and $\kappa^{\text{Li}}$ theoretically

The planar interface equilibrium densities can be computed using the Maxwell-rule which is equivalent to Eq. (12). This equation have three variables $\rho_V$, $\rho_l$ and $p_0$. To complete a set of three equations we also use $p_{EOS}(\rho_v) = p_0$ and $p_{EOS}(\rho_l) = p_0$.

Next, we want to make the P-LBM mechanical stability condition also satisfy this equation. In other worlds, we want to find $\epsilon$ that makes the integral in Eq. (6) be equal to zero for the same densities obtained from the Maxwell-rule. Then, we define:

$$F(\epsilon) = \int_{\rho_v}^{\rho_l} (p_{EOS} - p_0) \frac{\dot{\psi}}{\psi^{1+\epsilon}} d\rho, \tag{13}$$

from this definition we also compute:

$$\frac{dF}{d\epsilon} = \int_{\rho_v}^{\rho_l} (p_{EOS} - p_0) \frac{\dot{\psi} \ln(\psi)}{\psi^{1+\epsilon}} d\rho. \tag{14}$$

The integrals in Eqs. (13) and (14) can be computed numerically using the Simpson's rule. Using the Newton method we compute $\Delta\epsilon = -F/(dF/d\epsilon)$ and update $\epsilon = \epsilon + \Delta\epsilon$ until a desired tolerance is achieved. The initial guess was $\epsilon = 2$. Note that a value for $\epsilon$ is obtained for each temperature. From $\epsilon$ we compute the $\sigma^{Li}$ value.

To compute $\kappa^{\text{Li}}$, we theoretically compute the surface tension which is given by the mismatch between the pressure components $p_{xx}$ and $p_{yy}$ (considering a planar interface along x-direction):

$$\gamma = \int_{-\infty}^{\infty} (p_{xx} - p_{yy}) dx. \tag{15}$$

Replacing the pressure tensor defined in Eq. (1) into the above equation:

$$\gamma = \frac{Gc^4(\Delta t)^4}{6}(1 - \kappa^{\text{Li}}) \int_{-\infty}^{\infty} \psi \frac{d^2\psi}{dx^2} dx. \tag{16}$$

Next, consider the following relation:

$$\psi \frac{d^2\psi}{dx^2} = \frac{d}{dx}\left(\psi \frac{d\psi}{dx}\right) - \left(\frac{d\psi}{dx}\right)^2. \tag{17}$$

We replace the above equation into Eq. (16) and use the fact that $d\psi/dx = 0$ at the boundaries to obtain:

$$\gamma = -\frac{Gc^4(\Delta t)^4}{6}(1 - \kappa^{\text{Li}}) \int_{-\infty}^{\infty} \left(\frac{d\psi}{dx}\right)^2 dx = -\frac{Gc^4(\Delta t)^4}{6}(1 - \kappa^{\text{Li}}) \int_{\rho_v}^{\rho_l} \dot{\psi} \frac{d\psi}{dx} d\rho. \tag{18}$$

We converted the above integral in terms of $x$ into an integral in terms of $\rho$ by considering $(d\psi/dx)dx = (d\psi/d\rho)d\rho = \dot{\psi} d\rho$

Note that we can use Eq. (5) to compute $d\psi/dx$ as a function of $\rho$. Thus, we can numerically integrate (using the Simpson's rule) the right-hand-side of Eq. (18) to compute the surface tension. If we want to achieve a desired surface tension $\gamma$, we can explicitly calculate $\kappa^{\text{Li}}$ from the above equation.

## C. Modified pseudopotential method

We start performing a Champman-Enskog (Chapman and Cowling, 1990) analysis of the modified pseudopotential method given by Eqs. (1), (13) and (32) in the main paper. For convenience, we rewrite the lattice Boltzmann equation (LBE) in the following form:

$$f_i(t + \Delta t, \mathbf{x} + \mathbf{c}_i \Delta t) - f_i(t, \mathbf{x}) = \Delta t \Omega_i + \Delta t S_i + \Delta t C_i, \tag{19}$$

where:

$$\Omega_i = -[\mathbf{M}^{-1} \mathbf{\Lambda} \mathbf{M}]_{ij}(f_j - f_j^{eq}), \tag{20a}$$

$$S_i = \left[\mathbf{M}^{-1}\left(\mathbf{I} - \frac{\Lambda \Delta t}{2}\right)\mathbf{M}\right]_{ij} F'_j. \tag{20b}$$

The equilibrium distribution function moments are known:

$$\begin{aligned}
\sum_i f_i^{eq} &= \rho, \\
\sum_i c_{i\alpha} f_i^{eq} &= \rho u_\alpha, \\
\sum_i c_{i\alpha} c_{i\beta} f_i^{eq} &= \rho u_\alpha u_\beta + \rho c_s^2 \delta_{\alpha\beta}, \\
\sum_i c_{i\alpha} c_{i\beta} c_{i\gamma} f_i^{eq} &= \rho c_s^2 (u_\alpha \delta_{\beta\gamma} + u_\beta \delta_{\alpha\gamma} + u_\gamma \delta_{\alpha\beta}).
\end{aligned} \tag{21}$$

The forcing scheme moments are:

$$\begin{aligned}
\sum_i S_i &= 0, \\
\sum_i c_{i\alpha} S_i &= \left(1 - \frac{\omega_j \Delta t}{2}\right) F_\alpha, \\
\sum_i c_{i\alpha} c_{i\beta} S_i &= \left(1 - \frac{\omega_v \Delta t}{2}\right)(u_\alpha F_\beta + u_\beta F_\alpha) + \frac{\omega_v - \omega_e}{2} \Delta t \delta_{\alpha\beta} u_\gamma F_\gamma + 2\omega_e \frac{\kappa^P \sigma^{Li} |\mathbf{F}_{\text{int}}|^2}{\psi^2} \delta_{\alpha\beta}.
\end{aligned} \tag{22}$$

The relaxation frequencies $\omega$ are equal to the inverse of the relaxation times $\omega = 1/\tau$. The source term moments are:

$$\begin{aligned}
\sum_i C_i &= 0, \\
\sum_i c_{i\alpha} C_i &= 0, \\
\sum_i c_{i\alpha} c_{i\beta} C_i &= -\omega_v Q_{\alpha\beta} + \frac{\omega_v + 0.5\omega_e}{2} \delta_{\alpha\beta}(Q_{xx} + Q_{yy}).
\end{aligned} \tag{23}$$

The following distribution function moments are known:

$$\begin{aligned}
\sum_i f_i &= \rho, \\
\sum_i c_{i\alpha} f_i &= \rho u_\alpha - \frac{\Delta t F_\alpha}{2}.
\end{aligned} \tag{24}$$

The collision term is related with the distribution and equilibrium distribution functions, then, the following moments are known:

$$\begin{aligned}
\sum_i \Omega_i &= 0. \\
\sum_i c_{i\alpha} \Omega_i &= \frac{\omega_j \Delta t}{2} F_\alpha.
\end{aligned} \tag{25}$$

After expanding the LBE in taylor series, and performing some manipulations, we arrive in the form:

$$(\partial_t + c_{i\alpha}\partial_\alpha)f_i + \frac{\Delta t}{2}(\partial_t + c_{i\alpha}\partial_\alpha)(\Omega_i + S_i + C_i) = \Omega_i + S_i + C_i. \tag{26}$$

The next step is apply the following substitutions:

$$\begin{aligned}
\partial_t &= \epsilon \partial_t^{(1)} + \epsilon^2 \partial_t^{(2)}, \\
\partial_\alpha &= \epsilon \partial_\alpha^{(1)}, \\
f_i &= f_i^{eq} + \epsilon f_i^{(1)} + \epsilon^2 f_i^{(2)} + ..., \\
\Omega_i &= \epsilon \Omega_i^{(1)} + \epsilon^2 \Omega_i^{(2)} + ..., \\
S_i &= \epsilon S_i^{(1)}, \quad C_i = \epsilon C_i^{(1)}.
\end{aligned} \tag{27}$$

The following constraints are considered:

$$\begin{aligned}
&\sum_i f_i^{(1)} = 0, \quad \sum_i \Omega_i^{(k)} = 0, \quad \text{if} \quad k \geq 1, \\
&\sum_i c_{i\alpha} f_i^{(k)} = 0, \quad \sum_i c_{i\alpha} \Omega_i^{(k)} = 0, \quad \text{if} \quad k \geq 2. \\
&\sum_i c_{i\alpha} f_i^{(1)} = -\frac{\Delta t F_\alpha^{(1)}}{2}, \quad \sum_i c_{i\alpha} \Omega_i^{(1)} = \frac{\omega_j \Delta t}{2} F_\alpha^{(1)}.
\end{aligned} \tag{28}$$

The relation between $f_i$ and $\Omega_i$ also implies:

$$\frac{1}{\omega_v} \sum_i c_{i\alpha} c_{i\beta} \Omega_i^{(1)} + \left(\frac{1}{\omega_e} - \frac{1}{\omega_v}\right) \frac{\delta_{\alpha\beta}}{2} \sum_i c_{i\gamma} c_{i\gamma} \Omega_i^{(1)} \\
= - \sum_i c_{i\alpha} c_{i\beta} f_i^{(1)}. \tag{29}$$

Applying the expansion, we obtain the equations $O(\epsilon)$ and $O(\epsilon^2)$:

$$O(\epsilon): \quad \partial_t^{(1)} f_i^{eq} + c_{i\alpha} \partial_\alpha^{(1)} f_i^{eq} = \Omega_i^{(1)} + S_i^{(1)} + C_i^{(1)}, \tag{30a}$$

$$O(\epsilon^2): \quad \partial_t^{(2)} f_i^{eq} + (\partial_t^{(1)} + c_{i\alpha} \partial_\alpha^{(1)}) f_i^{(1)} + \frac{\Delta t}{2} (\partial_t^{(1)} + c_{i\alpha} \partial_\alpha^{(1)})(\Omega_i^{(1)} + S_i^{(1)} + C_i^{(1)}) = \Omega_i^{(2)}. \tag{30b}$$

With the knowledge of all previous moments we can compute the zeroth order moment ($M_0$) of the above equations ($O(\epsilon)$ and $O(\epsilon^2)$):

$$M_0(\epsilon): \quad \partial_t^{(1)} \rho + \partial_\alpha^{(1)}(\rho u_\alpha) = 0, \tag{31a}$$

$$M_0(\epsilon^2): \quad \partial_t^{(2)} \rho = 0. \tag{31b}$$

Combining both equations ($\epsilon M_0(\epsilon) + \epsilon^2 M_0(\epsilon^2)$), one recovers the mass conservation equation:

$$\partial_t \rho + \partial_\alpha(\rho u_\alpha) = 0. \tag{32}$$

Next, we compute the first order moments ($M_1$) of $O(\epsilon)$ and $O(\epsilon^2)$:

$$M_1(\epsilon): \quad \partial_t^{(1)}(\rho u_\beta) + \partial_\alpha^{(1)}(\rho u_\alpha u_\beta + \rho c_s^2 \delta_{\alpha\beta}) = F_\beta^{(1)}, \tag{33a}$$

$$M_1(\epsilon^2): \quad \partial_t^{(2)}(\rho u_\beta) + \partial_\alpha^{(1)} \left[\sum c_{i\alpha} c_{i\beta} \left(f_i^{(1)} + \frac{\Delta t}{2} \Omega_i^{(1)} + \frac{\Delta t}{2} S_i^{(1)} + \frac{\Delta t}{2} C_i^{(1)}\right)\right] = 0. \tag{33b}$$

There are two unknown moments in the above equation $\sum_i c_{i\alpha} c_{i\beta} f_i^{(1)}$ and $\sum_i c_{i\alpha} c_{i\beta} \Omega_i^{(1)}$. We use Eq. (29) to write the equation only in terms of the second term:

$$M_1(\epsilon^2): \quad \partial_t^{(2)}(\rho u_\beta) = \partial_\alpha^{(1)} \left[ \left( \frac{1}{\omega_\nu} - \frac{\Delta t}{2} \right) \sum c_{i\alpha} c_{i\beta} \Omega_i^{(1)} + \left( \frac{1}{\omega_e} - \frac{1}{\omega_\nu} \right) \frac{\delta_{\alpha\beta}}{2} \sum c_{i\gamma} c_{i\gamma} \Omega_i^{(1)} \right. \\ \left. - \frac{\Delta t}{2} \sum c_{i\alpha} c_{i\beta} S_i^{(1)} - \frac{\Delta t}{2} \sum c_{i\alpha} c_{i\beta} C_i^{(1)} \right]. \quad (34)$$

To complete analysis, we need an approximation for $\sum_i c_{i\alpha} c_{i\beta} \Omega_i^{(1)}$. We obtain this approximation by computing the second order moment ($M_2$) of $O(\epsilon)$:

$$\sum c_{i\alpha} c_{i\beta} \Omega_i^{(1)} + \sum c_{i\alpha} c_{i\beta} S_i^{(1)} + \sum c_{i\alpha} c_{i\beta} C_i^{(1)} = \partial_t^{(1)}(\rho u_\alpha u_\beta + \rho c_s^2) + \partial_\gamma^{(1)}[\rho c_s^2(u_\alpha \delta_{\beta\gamma} + u_\beta \delta_{\alpha\gamma} + u_\gamma \delta_{\alpha\beta})]. \quad (35)$$

After some length manipulations, the following result is obtained:

$$\sum c_{i\alpha} c_{i\beta} \Omega_i^{(1)} + \sum c_{i\alpha} c_{i\beta} S_i^{(1)} + \sum c_{i\alpha} c_{i\beta} C_i^{(1)} = u_\alpha F_\beta^{(1)} + u_\beta F_\alpha^{(1)} + \rho c_s^2 (\partial_\beta^{(1)} u_\alpha + \partial_\alpha^{(1)} u_\beta) \quad (36)$$

Replacing Eq. (36) and the moments defined in Eqs. (22) and (23) into Eq. (34):

$$M_1(\epsilon^2): \quad \partial_t^{(2)}(\rho u_\beta) = \partial_\alpha^{(1)} \left[ \left( \frac{1}{\omega_\nu} - \frac{\Delta t}{2} \right) \rho c_s^2 (\partial_\beta^{(1)} u_\alpha + \partial_\alpha^{(1)} u_\beta) + \left( \frac{1}{\omega_e} - \frac{1}{\omega_\nu} \right) \delta_{\alpha\beta} \rho c_s^2 (\partial_\gamma^{(1)} u_\gamma) \right. \\ \left. - 2 \frac{\sigma^{\text{Li}} |\mathbf{F}_{\text{int}}^{(1)}|^2}{\psi^2} \delta_{\alpha\beta} + Q_{\alpha\beta}^{(1)} - \frac{3}{4} Q_{\gamma\gamma}^{(1)} \delta_{\alpha\beta} \right]. \quad (37)$$

The final momentum conservation equation is obtained by the combination $\epsilon M_1(\epsilon) + \epsilon^2 M_1(\epsilon^2)$:

$$\partial_t(\rho u_\beta) + \partial_\alpha(\rho u_\alpha u_\beta) = -\partial_\alpha(\rho c_s^2 \delta_{\alpha\beta}) + F_\beta + \partial_\alpha \sigma'_{\alpha\beta} + \partial_\alpha \left( -2 \frac{\sigma^{\text{Li}} |\mathbf{F}_{\text{int}}|^2}{\psi^2} \delta_{\alpha\beta} + Q_{\alpha\beta} - \frac{3}{4} Q_{\gamma\gamma} \delta_{\alpha\beta} \right), \quad (38)$$

where $\sigma'_{\alpha\beta}$ is given by Eq. (8) in the main paper.

To complete the previous analysis is necessary to take into account the third order errors of the forcing scheme (Lycett-Brown and Luo, 2015) that are added to the right-hand-side of the momentum equation:

$$E_\alpha^{3rd} = \frac{c_s^2(\Delta t)^2}{12} \partial_\beta [(\partial_\gamma F_\gamma) \delta_{\alpha\beta} + \partial_\alpha F_\beta + \partial_\beta F_\alpha]. \quad (39)$$

To proceed with the analysis, we consider the following relations (Czelusniak, Mapelli, Guzella, Cabezas-Gómez and Wagner, 2020):

$$\sum_i \omega(|\mathbf{c}_i|^2) \phi(\mathbf{x} + \mathbf{c}_i \Delta t) = 3\phi(\mathbf{x}) + \frac{c^2 \Delta t^2}{2} \partial_\gamma \partial_\gamma \phi(\mathbf{x}) + O(c^4 \Delta t^4),$$

$$\sum_i \omega(|\mathbf{c}_i|^2) c_{i\alpha} \phi(\mathbf{x} + \mathbf{c}_i \Delta t) = c^2 \Delta t \partial_\alpha \phi(\mathbf{x}) + \frac{c^4 \Delta t^3}{2} \partial_\alpha \partial_\gamma \partial_\gamma \phi(\mathbf{x}) + O(c^6 \Delta t^5), \quad (40)$$

$$\sum_i \omega(|\mathbf{c}_i|^2)(c_{i\alpha} c_{i\beta} - c_s^2 \delta_{\alpha\beta}) \phi(\mathbf{x} + \mathbf{c}_i \Delta t) = \frac{c^4 \Delta t^2}{3} \partial_\alpha \partial_\beta \phi(\mathbf{x}) + O(c^6 \Delta t)^4.$$

Then, we apply these relations to determine continuous forms for the Shan-Chen force $\mathbf{F}_{\text{int}} = \mathbf{F}^{\text{SC}}$ (Eq. (10) in the main paper) and $Q_{\alpha\beta}$ (Eq. (30) in the main paper):

$$F_\alpha^{SC} = -Gc^2 \Delta t^2 \psi \partial_\alpha \psi - G \frac{c^4 \Delta t^4}{2} \psi \partial_\alpha \partial_\gamma \partial_\gamma \psi, \quad (41a)$$

$$Q_{\alpha\beta} = (1 - \kappa^P + \kappa^{\text{Li}}) \frac{Gc^4 (\Delta t)^4}{6} \psi \partial_\alpha \partial_\beta \psi + (5\kappa^P + \kappa^{\text{Li}} - 5) \frac{Gc^4 (\Delta t)^4}{12} \psi \partial_\gamma \partial_\gamma \psi \delta_{\alpha\beta}. \quad (41b)$$

With these information, we can compute the following terms:

$$2\frac{\sigma^{\text{Li}}|\mathbf{F}_{\text{int}}|^2}{\psi^2}\delta_{\alpha\beta} = 2G^2 c^4 \Delta t^4 \sigma^{\text{Li}}(\partial_\gamma \psi)(\partial_\gamma \psi), \tag{42a}$$

$$Q_{\alpha\beta} - \frac{3}{4}Q_{\gamma\gamma}\delta_{\alpha\beta} = (1 - \kappa^{\text{P}} + \kappa^{\text{Li}})\frac{Gc^4(\Delta t)^4}{6}\psi\partial_\alpha\partial_\beta\psi + (1 - \kappa^{\text{P}} - 2\kappa^{\text{Li}})\frac{Gc^4(\Delta t)^4}{12}\psi\partial_\gamma\partial_\gamma\psi\delta_{\alpha\beta}. \tag{42b}$$

$$\begin{aligned}-\partial_\alpha(\rho c_s^2 \delta_{\alpha\beta}) + F_\beta^{SC} + E_\beta^{3rd} = \\ \left(p_{EOS} + \frac{Gc^4(\Delta t)^4}{12}24G\kappa^{\text{P}}\sigma^{\text{Li}}(\partial_\gamma\psi)(\partial_\gamma\psi) + \frac{Gc^4(\Delta t)^4}{12}(\kappa^{\text{P}} + 2\kappa^{\text{Li}})\psi\partial_\gamma\partial_\gamma\psi\right)\delta_{\alpha\beta} \\ + \frac{Gc^4(\Delta t)^4}{6}(\kappa^{\text{P}} - \kappa^{\text{Li}})\psi\partial_\alpha\partial_\beta\psi.\end{aligned} \tag{42c}$$

Finally we combine all the terms of the right-hand-side of Eq. (38), with the exception of the viscous stress tensor and external forces, into a single pressure tensor to obtain the final form for the modified pseudopotential method (MP-LBM). The MP-LBM pressure tensor is given by:

$$\begin{aligned}p_{\alpha\beta} = \left(p_{EOS} + \frac{Gc^4(\Delta t)^4}{12}24G\kappa^{\text{P}}\sigma^{\text{Li}}(\partial_\gamma\psi)(\partial_\gamma\psi) + \frac{Gc^4(\Delta t)^4}{12}(\kappa^{\text{P}} + 2\kappa^{\text{Li}})\psi\partial_\gamma\partial_\gamma\psi\right)\delta_{\alpha\beta} \\ + \frac{Gc^4(\Delta t)^4}{6}(\kappa^{\text{P}} - \kappa^{\text{Li}})\psi\partial_\alpha\partial_\beta\psi.\end{aligned} \tag{43}$$

Comparing with Eqs. (1) and (2) we see that the new coefficients $A$ and $B$ are $A = 24G\kappa^{\text{P}}\sigma^{\text{Li}}$ and $B = 3\kappa^{\text{Li}}$. Which results in $\epsilon = -2A/B = -16G\sigma^{\text{Li}}$.

Then, the density profile is given by:

$$\left(\frac{d\psi}{dx}\right)^2 = (\dot\psi)^2\left(\frac{d\rho}{dx}\right)^2 = \frac{8\psi^\epsilon}{Gc^4(\Delta t)^4}\int_{\rho_v}^\rho \frac{(p_0 - p_{EOS})}{\kappa^{\text{P}}}\frac{\dot\psi}{\psi^{1+\epsilon}}d\rho, \tag{44}$$

which is dependent on $\kappa^{\text{P}}$ that can be adjusted to change the interface width.

The mechanical stability condition is unchanged:

$$\int_{\rho_v}^{\rho_l}(p_{EOS} - p_0)\frac{\dot\psi}{\psi^{1+\epsilon}}d\rho = 0, \tag{45}$$

with $\epsilon$ and $\sigma^{\text{Li}}$ maintaining the same relation.

Finally, the surface tension for the MP-LBM is given by:

$$\gamma = -\frac{Gc^4(\Delta t)^4}{6}(\kappa^{\text{P}} - \kappa^{\text{Li}})\int_{\rho_v}^{\rho_l}\dot\psi\frac{d\psi}{dx}d\rho. \tag{46}$$

Now, the surface tension depends on both $\kappa^{\text{P}}$ and $\kappa^{\text{Li}}$. In practice, we first adjust $\kappa^{\text{P}}$ to obtain the desired interface width and then we adjust $\kappa^{\text{Li}}$ to obtain the desired surface tension.

### D. Discrete planar interface equation for improved free energy

In this section, we will obtain the FE-LBM solution for a planar interface following a different procedure from Appendix A. Here we will obtain a discrete equation in terms of the macroscopic quantities (Peng, Ayala, Wang and Ayala, 2020; Guo, 2021).

In the planar interface problem, two phases are in equilibrium separated by an interface normal to the x-axis. In equilibrium and considering gradients only in the x-axis, we can drop the time and y-axis coordinates because $f_i(t + \Delta t, x, y) = f_i(t, x, y)$ and $f_i(t, x, y + \Delta y) = f_i(t, x, y)$. Considering these simplifying assumptions the MRT equation for the improved well balanced FE-LBM is:

$$f_i(x + c_{ix}\Delta t) = f_i(\Delta x) - \Delta t M_{ik}^{-1} \lambda_k m_k^{neq} + \Delta t M_{ik}^{-1} \left(1 - \frac{\lambda_k \Delta t}{2}\right) mF_k. \quad (47)$$

Where $m_k^{neq} = m_k - m_k^{eq}$ are the non-equilibrium moments and $mF_k = M_{kj}F'_j$ are the forcing moments.

The conversion matrix is given by:

$$\mathbf{M} = \begin{pmatrix} 1 & 1 & 1 & 1 & 1 & 1 & 1 & 1 & 1 \\ -4 & -1 & -1 & -1 & -1 & 2 & 2 & 2 & 2 \\ 4 & -2 & -2 & -2 & -2 & 1 & 1 & 1 & 1 \\ 0 & 1 & 0 & -1 & 0 & 1 & -1 & -1 & 1 \\ 0 & -2 & 0 & 2 & 0 & 1 & -1 & -1 & 1 \\ 0 & 0 & 1 & 0 & -1 & 1 & 1 & -1 & -1 \\ 0 & 0 & -2 & 0 & 2 & 1 & 1 & -1 & -1 \\ 0 & 1 & -1 & 1 & -1 & 0 & 0 & 0 & 0 \\ 0 & 0 & 0 & 0 & 0 & 1 & -1 & 1 & -1 \end{pmatrix}. \quad (48)$$

And it's inverse is:

$$\mathbf{M^{-1}} = \begin{pmatrix} 1/9 & -1/9 & 1/9 & 0 & 0 & 0 & 0 & 0 & 0 \\ 1/9 & -1/36 & -1/18 & 1/6 & -1/6 & 0 & 0 & 1/4 & 0 \\ 1/9 & -1/36 & -1/18 & 0 & 0 & 1/6 & -1/6 & -1/4 & 0 \\ 1/9 & -1/36 & -1/18 & -1/6 & 1/6 & 0 & 0 & 1/4 & 0 \\ 1/9 & -1/36 & -1/18 & 0 & 0 & -1/6 & 1/6 & -1/4 & 0 \\ 1/9 & 1/18 & 1/36 & 1/6 & 1/12 & 1/6 & 1/12 & 0 & 1/4 \\ 1/9 & 1/18 & 1/36 & -1/6 & -1/12 & 1/6 & 1/12 & 0 & -1/4 \\ 1/9 & 1/18 & 1/36 & -1/6 & -1/12 & -1/6 & -1/12 & 0 & 1/4 \\ 1/9 & 1/18 & 1/36 & 1/6 & 1/12 & -1/6 & -1/12 & 0 & -1/4 \end{pmatrix}. \quad (49)$$

There is a very important problem in the above definition of the conversion matrix. We are not providing the correct units of each term. Since the conversion functions are moments of velocity, it's components should have appropriate units. We will correct this issue by factoring the equilibrium and forcing moments. Then, we will guarantee that the distribution function have the correct units.

For this one-dimensional (1D) case $u_y = 0$, $F_y = 0$ and we define $u_x = U$ and $F_x = F$. The equilibrium distribution function and forcing moments are:

$$\mathbf{m^{eq}} = \begin{pmatrix} m_\rho^{eq} \\ m_e^{eq} \\ m_\xi^{eq} \\ m_{jx}^{eq} \\ m_{qx}^{eq} \\ m_{jy}^{eq} \\ m_{qy}^{eq} \\ m_v^{eq} \\ m_{pxy}^{eq} \end{pmatrix} = \begin{pmatrix} \rho \\ \rho(-4 + 3U^2 + 2A\Delta t \delta_x U)/c^2 \\ \rho(4 - 3U^2)/c^2 \\ \rho U/c \\ -\rho U/c \\ 0 \\ 0 \\ \rho U^2/c^2 \\ 0 \end{pmatrix}; \quad \mathbf{mF} = \begin{pmatrix} 0 \\ (6UF + 12c_s^2 U \delta_x \rho)/c^2 \\ (-6UF - 12c_s^2 U \delta_x \rho)/c^2 \\ F/c \\ -F/c \\ 0 \\ 0 \\ (2UF + 4c_s^2 U \Delta_x \rho)/c^2 \\ 0 \end{pmatrix}; \quad (50)$$

where we used the lattice speed $c = \Delta x/\Delta t$ to obtain consistent units. Note that these moments are consistent with Eqs (20) and (21), where we only neglected the terms related with the gravitational force. Also, we will consider $U = 0$ for our equilibrium planar interface system. The symbol $\delta_x$ is defined in Eq. (24) in the main paper.

For $i = 0, 2, 4$ the LBM equation is even simpler than Eq. (47) because $c_{0x} = c_{2x} = c_{4x} = 0$. For $i = 0$ and $i = 2$ the equations are:

$$M_{0k}^{-1} \lambda_k m_k^{neq} = M_{0k}^{-1} \left(1 - \frac{\lambda_k \Delta t}{2}\right) mF_k, \quad (51)$$

$$M_{2k}^{-1} \lambda_k m_k^{neq} = M_{2k}^{-1} \left(1 - \frac{\lambda_k \Delta t}{2}\right) mF_k. \tag{52}$$

Performing the summation for all $k$ values:

$$\frac{1}{9}\lambda_\rho m_\rho^{neq} - \frac{1}{9}\lambda_e m_e^{neq} + \frac{1}{9}\lambda_\xi m_\xi^{neq} = 0. \tag{53}$$

$$\frac{1}{9}\lambda_\rho m_\rho^{neq} - \frac{1}{36}\lambda_e m_e^{neq} - \frac{1}{18}\lambda_\xi m_\xi^{neq} - \frac{1}{4}\lambda_\nu m_\nu^{neq} = 0 \tag{54}$$

Eq. (53) is equivalent to Eq. (51) where for each index $k$ we replaced the moments shown in Eq (50) ($U = 0$). The same holds for Eqs (54) and (52). If we sum Eq. (53) with two times (2×) Eq. (54):

$$-\frac{1}{12}\lambda_e m_e^{neq} - \frac{1}{4}\lambda_\nu m_\nu^{neq} = 0. \tag{55}$$

In the above equation we considered the fact $m_\rho^{neq} = 0$. Now, we apply Eq. (47) for indexes $i = 1, 5, 8$ and sum all the equations:

$$f_1(x + \Delta x) + f_5(x + \Delta x) + f_8(x + \Delta x) - f_1(x) - f_5(x) - f_8(x) =$$
$$- \frac{\Delta t}{3}\lambda_\rho m_\rho^{neq} - \frac{\Delta t}{12}\lambda_e m_e^{neq} - \frac{\Delta t}{2}\lambda_j m_{jx}^{neq} - \frac{\Delta t}{4}\lambda_\nu m_\nu^{neq} + \frac{\Delta t}{2}\left(1 - \frac{\lambda_j \Delta t}{2}\right)\frac{F}{c}. \tag{56}$$

Where we know that $m_{jx}^{neq} = -F\Delta t/(2c)$. We can simply subtract Eq. (55) from Eq. (56) and obtain:

$$f_1(x + \Delta x) + f_5(x + \Delta x) + f_8(x + \Delta x) - f_1(x) - f_5(x) - f_8(x) = \frac{F\Delta t}{2c}. \tag{57}$$

We can write $f_1 + f_5 + f_8$ in the moment form $f_1 + f_5 + f_8 = m_\rho/3 + m_e/12 + m_{jx}/2 + m_\nu/4$. Also, we can write the unknown moments as $m_i = m_i^{eq} + m_i^{neq}$. Then, we obtain:

$$-\frac{F_{n+1}\Delta t}{4c} + \frac{m_{e;n+1}^{neq}}{12} + \frac{m_{\nu;n+1}^{neq}}{4} = +\frac{F_n \Delta t}{4c} + \frac{m_{e;n}^{neq}}{12} + \frac{m_{\nu;n}^{neq}}{4}. \tag{58}$$

The index $n + 1$ means that the variable is evaluated at position $x + \Delta x$ and $n$ simply evaluated at $x$. If we consider $\lambda_e = \lambda_\nu$ in Eq. (55) and replace the result in Eq. (58):

$$-\frac{F_{n+1}c\Delta t}{2} - \frac{F_n c\Delta t}{2} = 0 \rightarrow \frac{\rho_{n+1}\delta_x \mu_{n+1} + \rho_n \delta_x \mu_n}{2} = 0 \tag{59}$$

This equation implies $\mu = $ cte in equilibrium. We will define $\mu = \mu_0$. Then, $\mu_0 = \mu_b - \kappa \delta_x \delta_x \rho$. Here, the chemical potential is expressed in terms of $\delta_x$ because a finite difference stencil have to be used to approximate the derivative. Using Eq. (24) of the main paper as an approximation, we obtain:

$$\kappa \frac{(\rho_{i+1} - 2\rho_i + \rho_{i-1})}{\Delta x^2} = \mu_b - \mu_0. \tag{60}$$

Expanding in Taylor series, we have:

$$\kappa \frac{d^2\rho}{dx^2} + \kappa \frac{\Delta x^2}{12}\frac{d^4\rho}{dx^4} + O(\Delta x^4) = \mu_b - \mu_0. \tag{61}$$

Now, we will make the following definitions: $d\rho/dx = z$ and $z^2 = f_z$. Then, the following relations are valid:

$$\begin{aligned}\frac{d^2\rho}{dx^2} &= \dot{z}z = \frac{\dot{f}_z}{2}, \\ \frac{d^4\rho}{dx^4} &= \frac{\ddot{f}_z f_z}{2} + \frac{\dot{f}_z \dot{f}_z}{4} = \frac{1}{2}\frac{d}{d\rho}\left(\ddot{f}_z f_z - \frac{\dot{f}_z \dot{f}_z}{4}\right).\end{aligned} \tag{62}$$

Replacing these relation in Eq. (61) and performing some manipulations:

$$f_z\Big|_{\rho_v}^{\rho_l} + \frac{\Delta x^2}{12}\left(\ddot{f}_z f_z - \frac{\dot{f}_z \dot{f}_z}{4}\right)\Big|_{\rho_v}^{\rho_l} + O(\Delta x^4) = \int_{\rho_v}^{\rho_l} \frac{2(\mu_b - \mu_0)}{\kappa} d\rho. \tag{63}$$

We assume that the derivatives $\dot{z}$ and $\ddot{f}_z$ are bounded. Then, $\ddot{f}_z f_z = 0$ ($f_z = 0$) and $\dot{f}_z = 2z\dot{z} = 0$ ($z = 0$) at the boundaries. Applying the thermodynamic relations for the chemical potential, then, we can write $\int_{\rho_v}^{\rho_l}(\mu_b - \mu_0)d\rho = \int_{\rho_v}^{\rho_l}[(p_{EOS} - p_0)/\rho^2]d\rho$. Then, the final equation is:

$$\int_{\rho_v}^{\rho_l}(p_{EOS} - p_0)\frac{d\rho}{\rho^2} = O(\Delta x^4). \tag{64}$$

Which means that Eq. (12) is an approximation of at least order $O(\Delta x^4)$.

### E. Discrete planar interface equation for pseudopotential method

In this section, we will obtain the P-LBM discrete solution equation for a planar interface in terms of macroscopic quantities. The considerations are the same done in the previous section. The MRT equation for the MP-LBM is:

$$f_i(x + c_{ix}\Delta t) = f_i(\Delta x) - \Delta t M_{ik}^{-1}\lambda_k m_k^{neq} + \Delta t M_{ik}^{-1}\left(1 - \frac{\lambda_k \Delta t}{2}\right)mF_k + \Delta t M_{ik}^{-1}mC_k. \tag{65}$$

Where $m_k^{neq} = m_k - m_k^{eq}$ are the non-equilibrium moments, $mF_k = M_{kj}F'_j$ are the forcing moments and $mC_k = M_{kj}C_j$ are the source term moments. The conversion matrices are the same presented in the previous section.

The equilibrium distribution function, forcing and source terms moments are:

$$\mathbf{m^{eq}} = \begin{pmatrix} \rho \\ \rho(-2 + 3U^2/c^2) \\ \rho(1 - 3U^2/c^2) \\ \rho U/c \\ -\rho U/c \\ 0 \\ 0 \\ \rho U^2/c^2 \\ 0 \end{pmatrix}; \quad \mathbf{mF} = \begin{pmatrix} 0 \\ 6UF/c^2 + \frac{12\kappa^P \sigma^{Li} F^2}{\psi^2(\tau_e - \Delta t/2)c^2} \\ -6UF/c^2 - \frac{12\kappa^P \sigma^{Li} F^2}{\psi^2(\tau_\xi - \Delta t/2)c^2} \\ F/c \\ -F/c \\ 0 \\ 0 \\ 2UF/c^2 \\ 0 \end{pmatrix}; \quad \mathbf{mC} = \begin{pmatrix} 0 \\ 1.5\tau_e^{-1}(Q_{xx} + Q_{yy}) \\ -1.5\tau_\xi^{-1}(Q_{xx} + Q_{yy}) \\ 0 \\ 0 \\ 0 \\ 0 \\ -\tau_v^{-1}(Q_{xx} - Q_{yy}) \\ 0 \end{pmatrix}; \tag{66}$$

where we used the lattice speed $c = \Delta x/\Delta t$ to obtain consistent units. Again, we considered $u_y = 0$, $F_y = 0$, $u_x = U$ and $F_x = F$. In the next calculations we also consider $U = 0$ for a planar interface in equilibrium.

For $i = 0, 2, 4$ the LBM equation is even simpler than Eq. (65) because $c_{0x} = c_{2x} = c_{4x} = 0$. For $i = 0$ and $i = 2$ the equations are:

$$M_{0k}^{-1}\lambda_k m_k^{neq} = M_{0k}^{-1}\left(1 - \frac{\lambda_k \Delta t}{2}\right)mF_k + M_{0k}^{-1}mC_k, \tag{67}$$

$$M_{2k}^{-1}\lambda_k m_k^{neq} = M_{2k}^{-1}\left(1 - \frac{\lambda_k \Delta t}{2}\right)mF_k + M_{2k}^{-1}mC_k. \tag{68}$$

Performing the summation for all $k$ values:

$$\frac{1}{9}\lambda_\rho m_\rho^{neq} - \frac{1}{9}\lambda_e m_e^{neq} + \frac{1}{9}\lambda_\xi m_\xi^{neq} = -\frac{1}{9}\left(1 - \frac{\lambda_e \Delta t}{2}\right)\frac{12\kappa^P \sigma^{Li} F^2}{\psi^2(\tau_e - \Delta t/2)c^2} \\ - \frac{1}{9}\left(1 - \frac{\lambda_\xi \Delta t}{2}\right)\frac{12\kappa^P \sigma^{Li} F^2}{\psi^2(\tau_\xi - \Delta t/2)c^2} - \frac{1}{6}\lambda_e(Q_{xx} + Q_{yy}) - \frac{1}{6}\lambda_\xi(Q_{xx} + Q_{yy}). \quad (69)$$

$$\frac{1}{9}\lambda_\rho m_\rho^{neq} - \frac{1}{36}\lambda_e m_e^{neq} - \frac{1}{18}\lambda_\xi m_\xi^{neq} - \frac{1}{4}\lambda_\nu m_\nu^{neq} = -\frac{1}{36}\left(1 - \frac{\lambda_e \Delta t}{2}\right)\frac{12\kappa^P \sigma^{Li} F^2}{\psi^2(\tau_e - \Delta t/2)c^2} \\ + \frac{1}{18}\left(1 - \frac{\lambda_\xi \Delta t}{2}\right)\frac{12\kappa^P \sigma^{Li} F^2}{\psi^2(\tau_\xi - \Delta t/2)c^2} - \frac{1}{24}\lambda_e(Q_{xx} + Q_{yy}) + \frac{1}{12}\lambda_\xi(Q_{xx} + Q_{yy}) + \frac{1}{4}\lambda_\nu(Q_{xx} - Q_{yy}), \quad (70)$$

Eq. (69) is equivalent to Eq. (67) where for each index $k$ we replaced the moments shown in Eq (66). The same holds for Eqs (70) and (68). If we sum Eq. (69) with two times (2×) Eq. (70):

$$-\frac{1}{12}\lambda_e m_e^{neq} - \frac{1}{4}\lambda_\nu m_\nu^{neq} = -\frac{1}{12}\left(1 - \frac{\lambda_e \Delta t}{2}\right)\frac{12\kappa^P \sigma^{Li} F^2}{\psi^2(\tau_e - \Delta t/2)c^2} - \frac{1}{8}\lambda_e(Q_{xx} + Q_{yy}) + \frac{1}{4}\lambda_\nu(Q_{xx} - Q_{yy}). \quad (71)$$

In the above equation we considered the fact $m_\rho^{neq} = 0$.

Now, we apply Eq. (65) for indexes $i = 1, 5, 8$ and sum all the equations:

$$f_1(x + \Delta x) + f_5(x + \Delta x) + f_8(x + \Delta x) \\ - f_1(x) - f_5(x) - f_8(x) = \\ - \frac{\Delta t}{3}\lambda_\rho m_\rho^{neq} - \frac{\Delta t}{12}\lambda_e m_e^{neq} - \frac{\Delta t}{2}\lambda_j m_{jx}^{neq} - \frac{\Delta t}{4}\lambda_\nu m_\nu^{neq} \\ + \frac{\Delta t}{12}\left(1 - \frac{\lambda_e \Delta t}{2}\right)\frac{12\kappa^P \sigma^{Li} F^2}{\psi^2(\tau_e - \Delta t/2)c^2} \\ + \frac{\Delta t}{2}\left(1 - \frac{\lambda_j \Delta t}{2}\right)\frac{F}{c} + \frac{\Delta t}{8}\lambda_e(Q_{xx} + Q_{yy}) \\ - \frac{\Delta t}{4}\lambda_\nu(Q_{xx} - Q_{yy}). \quad (72)$$

Where we know that $m_{jx}^{neq} = -F\Delta t/(2c)$. We can simply subtract Eq. (71) from Eq. (72) and obtain:

$$f_1(x + \Delta x) + f_5(x + \Delta x) + f_8(x + \Delta x) - f_1(x) - f_5(x) - f_8(x) = \frac{F\Delta t}{2c}. \quad (73)$$

We can write $f_1 + f_5 + f_8$ in the moment form $f_1 + f_5 + f_8 = m_\rho/3 + m_e/12 + m_{jx}/2 + m_\nu/4$. Also, we can write the unknown moments as $m_i = m_i^{eq} + m_i^{neq}$. Then, we obtain:

$$\frac{\rho_{n+1}}{6} - \frac{F_{n+1}\Delta t}{4c} + \frac{m_{e;n+1}^{neq}}{12} + \frac{m_{\nu;n+1}^{neq}}{4} = \frac{\rho_n}{6} + \frac{F_n \Delta t}{4c} + \frac{m_{e;n}^{neq}}{12} + \frac{m_{\nu;n}^{neq}}{4}. \quad (74)$$

The index $n+1$ means that the variable is evaluated at position $x + \Delta x$ and $n$ simply evaluated at $x$. If we consider $\lambda_e = \lambda_\nu$ in Eq. (71) and replace the result in Eq. (74):

$$\frac{c^2}{3}\rho_{n+1} - \frac{F_{n+1}c\Delta t}{2} + 2\frac{\kappa^P \sigma^{Li} F_{n+1}^2}{\psi_{n+1}^2} - \frac{Q_{xx;n+1}}{4} + \frac{3Q_{yy;n+1}}{4} = \frac{c^2}{3}\rho_n + \frac{F_n c\Delta t}{2} + 2\frac{\kappa^P \sigma^{Li} F_n^2}{\psi_n^2} - \frac{Q_{xx;n}}{4} + \frac{3Q_{yy;n}}{4} \quad (75)$$

To obtain a final form for the discrete macroscopic planar interface equation we just need to replace the expressions for $F$, $Q_{xx}$, $Q_{yy}$ and perform some tedious manipulations. Then, we arrive in the following expression:

$$c_s^2 \rho_n + \frac{Gc^2(\Delta t)^2}{2}\psi_n^2 + \frac{Gc^4(\Delta t)^4}{12}3\kappa^P\psi_n\frac{\psi_{n+1} - 2\psi_n + \psi_{n-1}}{(\Delta x)^2} + \frac{Gc^4(\Delta t)^4}{12}24G\kappa^P\sigma^{\text{Li}}\left(\frac{\psi_{n+1} - \psi_{n-1}}{2\Delta x}\right)^2 = \text{cte}. \quad (76)$$

Looking to Eq. (11) in the main paper, we arrive in the conclusion that the first two terms of Eq. (76) are equivalent to $p_{EOS}$. In the bulk phases, $\psi$ is approximately constant, then cte is equal to the bulk pressure $p_0$.

Before doing an error analysis of Eq. (76) we must notice that $\psi$ also depends on $\Delta x$ in its definition, Eq (11). To eliminate this dependency, we define $\bar{\psi} = \Delta x \psi$. Then, the equation becomes:

$$p_{EOS} + \frac{G(\Delta x)^2}{12}3\kappa^P\bar{\psi}_n\frac{\bar{\psi}_{n+1} - 2\bar{\psi}_n + \bar{\psi}_{n-1}}{(\Delta x)^2} + \frac{G(\Delta x)^2}{12}24G\kappa^P\sigma^{\text{Li}}\left(\frac{\bar{\psi}_{n+1} - \bar{\psi}_{n-1}}{2\Delta x}\right)^2 = p_0, \quad (77)$$

where $c\Delta t = \Delta x$ was used. Before continuing, let us rewrite Eq. (2) using the $\bar{\psi}$ definition:

$$p_{EOS} + \frac{G(\Delta x)^2}{12}3\kappa^P\bar{\psi}\frac{d^2\bar{\psi}}{dx^2} + \frac{G(\Delta x)^2}{12}24G\kappa^P\sigma^{\text{Li}}\left(\frac{d\bar{\psi}}{dx}\right)^2 = p_0. \quad (78)$$

We see that Eq. (77) is the discrete version of Eq. (78). From the last one, we obtained the mechanical stability solution Eq. (6). However in the previous analysis we were not taking into account the effect of the discretization errors.

So, our intention here is to compare the solution of Eq. (77) with Eq. (78) and see the effect of $\Delta x$ as we did for the FE-LBM. However, we have a problem in this case. Even our target continuous equation - Eq. (78) - depends on the value of $\Delta x$. To correct this issue, we define $\bar{\kappa}^P = (\Delta x)^2 \kappa^P$. Then, we obtain a pressure tensor in physical units that is independent on the grid size.

Now, the discrete equation can be written as:

$$\bar{A}\left(\frac{\bar{\psi}_{n+1} - \bar{\psi}_{n-1}}{2\Delta x}\right)^2 + \bar{B}\bar{\psi}_n\frac{\bar{\psi}_{n+1} - 2\bar{\psi}_n + \bar{\psi}_{n-1}}{(\Delta x)^2} = p_0 - p_{EOS}, \quad (79)$$

where:

$$\bar{A} = \frac{G}{12}24G\bar{\kappa}^P\sigma^{\text{Li}},$$
$$\bar{B} = \frac{G}{12}3\bar{\kappa}^P. \quad (80)$$

Using the definition $\epsilon = -2\bar{A}/\bar{B}$ and applying the Taylor series expansion:

$$-\frac{\epsilon}{2}\left(\frac{d\bar{\psi}}{dx}\right)^2 + \bar{\psi}\frac{d^2\bar{\psi}}{dx^2} - \frac{\epsilon\Delta x^2}{6}\frac{d\bar{\psi}}{dx}\frac{d^3\bar{\psi}}{dx^3} + \frac{\Delta x^2}{12}\bar{\psi}\frac{d^4\bar{\psi}}{dx^4} + O(\Delta x^4) = \frac{p_0 - p_{EOS}}{\bar{B}}. \quad (81)$$

Using the definitions $d\bar{\psi}/dx = z$ and $z^2 = f_z$, we can write the following equations:

$$\begin{aligned}
\frac{d\bar{\psi}}{dx} &= z, \\
\frac{d^2\bar{\psi}}{dx^2} &= \frac{dz}{d\bar{\psi}}\frac{d\bar{\psi}}{dx} = \dot{z}z = \frac{\dot{f}_z}{2}, \\
\frac{d^3\bar{\psi}}{dx^3} &= \frac{\ddot{f}_z z}{2}, \\
\frac{d^4\bar{\psi}}{dx^4} &== \frac{\dddot{f}_z f_z}{2} + \frac{\ddot{f}_z \dot{f}_z}{4} = \frac{1}{2}\frac{d}{d\bar{\psi}}\left(\ddot{f}_z f_z - \frac{\dot{f}_z \dot{f}_z}{4}\right).
\end{aligned} \quad (82)$$

Replacing these expressions into Eq. (81) and doing some manipulations:

$$\frac{d}{d\bar{\psi}}\left(\frac{f_z}{\bar{\psi}^\epsilon}\right) - \frac{\epsilon \Delta x^2}{6} \frac{\ddot{f}_z f_z}{\bar{\psi}^{1+\epsilon}} + \frac{\Delta x^2}{12} \frac{1}{\bar{\psi}^\epsilon} \frac{d}{d\bar{\psi}}\left(\ddot{f}_z f_z - \frac{\dot{f}_z \dot{f}_z}{4}\right) + O(\Delta x^4) = \frac{2(p_0 - p_{EOS})}{\bar{B}\bar{\psi}^{1+\epsilon}}. \quad (83)$$

The mechanical stability condition can be obtained by simply integrating (in terms of $\bar{\psi}$) the above equation from the vapor phase $\bar{\psi}_v$ to the liquid phase $\bar{\psi}_l$.

We are going to solve the terms of the above integral separately. The first term, will be:

$$\int_{\bar{\psi}_v}^{\bar{\psi}_l} \frac{1}{\bar{\psi}^\epsilon} \frac{d}{d\bar{\psi}}\left(\ddot{f}_z f_z - \frac{\dot{f}_z \dot{f}_z}{4}\right) d\bar{\psi} = \frac{1}{\bar{\psi}^\epsilon}\left(\ddot{f}_z f_z - \frac{\dot{f}_z \dot{f}_z}{4}\right)\bigg|_{\bar{\psi}_v}^{\bar{\psi}_l \to 0} + \epsilon \int_{\bar{\psi}_v}^{\bar{\psi}_l} \frac{1}{\bar{\psi}^{1+\epsilon}}\left(\ddot{f}_z f_z - \frac{\dot{f}_z \dot{f}_z}{4}\right) d\bar{\psi}. \quad (84)$$

Assuming that $\dot{z}$ and $\dot{f}_z$ is bounded, we conclude that the first term of the right-hand side is equal to zero ($d\bar{\psi}/dx$ is zero at the boundaries). The other term is given by:

$$\int_{\bar{\psi}_v}^{\bar{\psi}_l} \frac{\ddot{f}_z f_z}{\bar{\psi}^{1+\epsilon}} d\bar{\psi} = \int_{\bar{\psi}_v}^{\bar{\psi}_l} \frac{1}{\bar{\psi}^{1+\epsilon}} \frac{d}{d\bar{\psi}}(\dot{f}_z f_z) d\bar{\psi} - \int_{\bar{\psi}_v}^{\bar{\psi}_l} \frac{\dot{f}_z \dot{f}_z}{\bar{\psi}^{1+\epsilon}} d\bar{\psi} \quad (85a)$$

$$\int_{\bar{\psi}_v}^{\bar{\psi}_l} \frac{1}{\bar{\psi}^{1+\epsilon}} \frac{d}{d\bar{\psi}}(\dot{f}_z f_z) d\bar{\psi} = \frac{\dot{f}_z f_z}{\bar{\psi}^{1+\epsilon}}\bigg|_{\bar{\psi}_v}^{\bar{\psi}_l \to 0} + \frac{(1+\epsilon)}{2} \int_{\bar{\psi}_v}^{\bar{\psi}_l} \frac{(df_z^2/d\bar{\psi})}{\bar{\psi}^{2+\epsilon}} d\bar{\psi} \quad (85b)$$

$$\int_{\bar{\psi}_v}^{\bar{\psi}_l} \frac{(df_z^2/d\bar{\psi})}{\bar{\psi}^{2+\epsilon}} d\bar{\psi} = \frac{f_z^2}{\bar{\psi}^{2+\epsilon}}\bigg|_{\bar{\psi}_v}^{\bar{\psi}_l \to 0} + (2+\epsilon) \int_{\bar{\psi}_v}^{\bar{\psi}_l} \frac{f_z^2}{\bar{\psi}^{3+\epsilon}} d\bar{\psi} \quad (85c)$$

Again, we consider $f_z = 0$ at the boundaries (absence of $\bar{\psi}$ gradients at bulk phases in equilibrium).

Adding these results in Eq. (83):

$$\frac{f_z}{\bar{\psi}^\epsilon}\bigg|_{\bar{\psi}_v}^{\bar{\psi}_l \to 0} - \frac{\epsilon(1+\epsilon)(2+\epsilon)}{24} \Delta x^2 \int_{\bar{\psi}_v}^{\bar{\psi}_l} \frac{f_z f_z}{\bar{\psi}^{3+\epsilon}} d\bar{\psi} + \frac{3\epsilon}{48} \Delta x^2 \int_{\bar{\psi}_v}^{\bar{\psi}_l} \frac{\dot{f}_z \dot{f}_z}{\bar{\psi}^{1+\epsilon}} d\bar{\psi} = \int_{\bar{\psi}_v}^{\bar{\psi}_l} \frac{2(p_0 - p_{EOS})}{\bar{B}\bar{\psi}^{1+\epsilon}} d\bar{\psi}. \quad (86)$$

The first thing to pay attention is that the integrals of the left-hand-side are non-negative (and also non-zero) since $\psi$, $f_z f_z$ and $\dot{f}_z \dot{f}_z$ are positive. Unless, both integrals cancel each other, which we cannot guarantee, the mechanical stability condition of Eq. (6) (Eq. (18) in the main paper) will not be true. The mechanical stability condition is then, writen as:

$$\int_{\rho_v}^{\rho_l} (p_{EOS} - p_0) \frac{\dot{\psi}}{\bar{\psi}^{1+\epsilon}} d\rho = O(\Delta x^2). \quad (87)$$

## F. Two-phase flow between parallel plates solution

First, we consider the discrete interface case. A steady state two-dimensional flow is confined between two parallel plates separated by a vertical distance $H$. The bottom part $0 \leq y \leq 0.5H$ is filled with fluid 1 and the top $0.5H < y \leq H$ with fluid 2. Gravity is neglected. The flow is driven by a force $F_x$ in the horizontal axis. Since we have a discontinuity in viscosity $\mu$ across the interface, we solve two equations. One equation for the fluid 1 velocity ($u_1$) and other for the fluid 2 velocity ($u_2$). Then, we apply the boundary conditions at the center line $y = H/2$: the two equations should provide the same velocity and shear stress at this position. Also, no-slip conditions at the walls are applied.

Since it is a well known result, we just provide the solution:

$$\begin{aligned}
C_1 &= \frac{F_x H}{4} \frac{\mu_2 - \mu_1}{\mu_1(\mu_1 + \mu_2)}, \\
C_2 &= -\frac{F_x H}{8\mu_1} - \frac{C_1 H}{2}, \\
u_1 &= -\frac{F_x}{2\mu_1}\left(\frac{H}{2} - y\right)^2 + C_1\left(\frac{H}{2} - y\right) + C_2, \\
u_2 &= -\frac{F_x}{2\mu_2}\left(y - \frac{H}{2}\right)^2 - \frac{\mu_1 C_1}{\mu_2}\left(y - \frac{H}{2}\right) + C_2.
\end{aligned} \quad (88)$$

For the diffuse interface case, instead of heaving a fluid 1 and 2 with different properties, we have only one fluid which properties change continuously. Then the viscosity is a function of the channel position $\mu = \mu(y)$. The solution is obtained by direct integration of the momentum conservation equation:

$$u(y) = -F_x \int_0^y \frac{y'}{\mu(y')} dy' + \mu(0)\frac{du}{dy}(0) \int_0^y \frac{1}{\mu(y')} dy' \quad (89)$$

where $\mu(0) = \mu(y = 0)$, the same is valid for $du/dy$. The viscosity $\mu$ is a function of the density in LBM where $\mu = \rho c_s^2 (\tau - 0.5)$. For a fixed $\tau$ we simple have $\mu \propto \rho$. The theoretical density profile was already derived in Appendix A.

## References

Chapman, S., Cowling, T.G., 1990. The mathematical theory of non-uniform gases: an account of the kinetic theory of viscosity, thermal conduction and diffusion in gases. Cambridge university press.

Czelusniak, L.E., Mapelli, V.P., Guzella, M., Cabezas-Gómez, L., Wagner, A.J., 2020. Force approach for the pseudopotential lattice boltzmann method. Physical Review E 102, 033307.

Guo, Z., 2021. Well-balanced lattice boltzmann model for two-phase systems. Physics of Fluids 33, 031709.

Krüger, T., Kusumaatmaja, H., Kuzmin, A., Shardt, O., Silva, G., Viggen, E.M., 2017. The lattice boltzmann method. Springer International Publishing 10, 4–15.

Li, Q., Luo, K., 2013. Achieving tunable surface tension in the pseudopotential lattice boltzmann modeling of multiphase flows. Physical Review E 88, 053307.

Lycett-Brown, D., Luo, K.H., 2015. Improved forcing scheme in pseudopotential lattice boltzmann methods for multiphase flow at arbitrarily high density ratios. Physical Review E 91, 023305.

Peng, C., Ayala, L.F., Wang, Z., Ayala, O.M., 2020. Attainment of rigorous thermodynamic consistency and surface tension in single-component pseudopotential lattice boltzmann models via a customized equation of state. Physical Review E 101, 063309.

Shan, X., 2008. Pressure tensor calculation in a class of nonideal gas lattice boltzmann models. Physical Review E 77, 066702.

Shan, X., Chen, H., 1994. Simulation of nonideal gases and liquid-gas phase transitions by the lattice boltzmann equation. Physical Review E 49, 2941.